\documentclass[preprint,12pt]{elsarticle}
\usepackage{amssymb}
\usepackage{amsmath}
\usepackage{bbold}
\usepackage{algorithm}
\usepackage{algpseudocode}
\usepackage{color}
\usepackage{lineno,hyperref}
\usepackage{subfigure}
\modulolinenumbers[5]
\newcounter{bla}

\newcommand {\ket}[1]{\ensuremath{| {#1} \rangle}}
\newcommand {\bra}[1]{\ensuremath{\langle {#1} |}}
\newcommand{\braket}[2]{\langle \ensuremath{#1} |\ensuremath{#2} \rangle}
\newcommand{\dd}{\mathrm{d}}
\begin{document}
\begin{frontmatter}
\title{GPU-accelerated algorithms for many-particle continuous-time
quantum walks}
\author[Bologna,Modena]{Enrico Piccinini\corref{correspondingauthor}}
\cortext[correspondingauthor]{Corresponding author.\\\textit{E-mail address:} enrico.piccinini@unimore.it}
\author[Milano]{Claudia Benedetti}
\author[Modena]{Ilaria Siloi}
\author[Milano,S3]{Matteo G. A. Paris}
\author[Modena,S3]{Paolo Bordone}
\address[Bologna]{Dipartimento di Ingegneria dell'Energia Elettrica e dell'Informazione 
``Guglielmo Marconi" - DEI, Universit\`a di Bologna,  I-40136 - Bologna, Italy }
\address[Milano]{Quantum Technology Lab, Dipartimento di Fisica, 
Universit\`a degli Studi di Milano, I-20133- Milano, Italy}
\address[Modena]{Dipartimento di Scienze Fisiche, Informatiche e Matematiche - 
FIM, Universit\`a di Modena e Reggio Emilia, I-41125 - Modena, Italy }
\address[S3]{Centro S3, CNR-Istituto di Nanoscienze,  I-41125 - Modena, Italy}
\begin{abstract}
Many-particle continuous-time quantum walks (CTQWs) represent a resource for several tasks in quantum
technology, including quantum search algorithms and universal quantum computation. In order to design and
implement CTQWs in a realistic scenario, one needs effective simulation tools for Hamiltonians that take into account static noise and fluctuations in the lattice, i.e. Hamiltonians containing stochastic terms.
To this aim, we suggest a parallel algorithm based on the 
Taylor series expansion of the evolution operator, and compare its performances with those of algorithms based on the exact diagonalization of the Hamiltonian or a 4-th order Runge-Kutta integration. We prove that both Taylor-series expansion and Runge-Kutta algorithms are reliable and have a low computational cost, the Taylor-series expansion showing the additional advantage of a memory allocation not depending on the precision of calculation. Both algorithms are also highly parallelizable within the SIMT paradigm, and are thus suitable for GPGPU computing. In turn, we have benchmarked 4 NVIDIA GPUs and 3 quad-core Intel CPUs for a 2-particle system over lattices of increasing dimension, showing that the speedup providend by GPU computing, with respect to the OPENMP parallelization, lies in the range between 8x and (more than) 20x, depending on the frequency of post-processing. GPU-accelerated codes thus allow one to overcome concerns about the execution time, and make it possible simulations with many interacting particles on large lattices, with the only limit of the memory available on the device.

\end{abstract}


\end{frontmatter}

\section{Introduction}
Quantum walks  (QWs) are a generalization of classical random walks to the quantum regime. They were first introduced  in  the discrete-time version~\cite{aharonov93} and later as continuous-time quantum walks (CTWQs) in the context of quantum computation and decision trees~\cite{fahri98}. In this framework, it has been shown that single-particle quantum walk-based algorithms may outperform the classical counterpart in terms of traveling time through a graph. Since then, QWs, both in the continuous- and discrete-time versions, have been the subject of extensive studies. Besides, QWs have been generalized to many-particles quantum walks, where the time evolution of the walkers depends upon their statistics, indistinguishability and kind of interaction~\cite{lahini12,benedetti12,wang14,Qin14}. CTQWs on more complex structures, e.g. complex graphs,  have been also the focus of more recent analysis~\cite{faccin13,caruso16,Schreiber55,hans16}. Overall, CTQWs have  been proved a useful tool in a variety of contexts,  ranging from  transport through  a graph~\cite{blumen11},  to quantum search algorithms~\cite{childs04, yasser16}, graph isomorphism testing~\cite{Wang08,gamble10,berry11} and universal quantum computation~\cite{childs09,Childs791}.

In realistic experimental scenarios, imperfections in the fabrication of the lattice may induce Anderson localization of the walkers~\cite{lahini10,schreiber11,crespi13,ghosh14}, whereas stochastic fluctuations of 
the environment may come into play destroying the quantumness of the system and, in turn, its peculiar propagation features~\cite{de14,benedetti16,siloi16,beggi16}.  A more realistic description for noisy quantum walks should therefore take into account  the noise that may affect the evolution of the walkers. A convenient way to describe noise is to introduce suitable stochastic terms in the  Hamiltonian, in order to model static or dynamical fluctuations that may affect both the on-site  energies  or the tunneling amplitudes of the walkers~\cite{lee14,benedetti16,siloi16}. The dynamical evolution of the QW is then obtained as the ensemble average over all possible realizations of the stochastic processes mimicking the noise. In practice, the ensemble average is computed numerically as an average over a finite number of realizations: the larger the number of the realizations, the more accurate the simulation of the CTQW. 

Evaluating the dynamics of a many-particle state over a noisy lattice requires the numerical solution of a set of differential equations which include stochastic terms~\cite{Hanggi82}. The total number of equations to solve grows rapidly as long as the numbers of nodes, particles, and realizations increase, thus making the problem more and more computationally demanding with longer execution times. In fact, codes for simulating many-particle CTQWs have been developed for high-performance clusters with distributed memory~\cite{Izaac15}. On the other hand, the evolution of computer architectures towards multicore processors even in stand-alone workstations enabled important cuts of the execution time by introducing the possibility of running multiple threads in parallel and spreading the workload among cores. This possibility was boosted up by the general purpose parallel computing architectures of modern graphic cards (GPGPUs). In the latter, hundreds or thousands of computational cores in the same single chip are able to process simultaneously a very large number of data. It should also be noted that an impressive computational power is present not only in dedicated GPUs for high-performance computing, but also in commodity graphic cards, which make modern workstations suitable for numerical analyses. In order to exploit such a huge computational power, algorithms must be first redesigned and adapted to the SIMT (Single Instruction Multiple Thread) and SIMD (Single Instruction Multiple Data) paradigms and translated then into programming languages with hardware-specific subsets of instructions. Among them, one of the most diffuse is CUDA-C, a C extension for the Compute Unified Device Architecture (CUDA) that represents the core component of NVIDIA GPUs. As a matter of fact, the use of GPUs for scientific analysis, which dates back to mid and late 2000's~\cite{Anderson07,Anderson08,Tolke08, Preis09,Januszewski10}, dramatically boosted with a two-digit yearly increasing rate since 2010.  Just looking at the computational physics realm, several GPU-specific algorithms have been proposed in the last three years, e.g. for stochastic differential equations~\cite{Spiechowicz15}, molecular dynamics simulations~\cite{Glaser15}, fluid dynamics \cite{Smith13, Januszewski14}, Metropolis Monte Carlo~\cite{Anderson16} simulations, quantum Monte Carlo simulations~\cite{Lutsyshyn15}, and free-energy calculations~\cite{Januszewski15}.


The evolution of many particles QWs in a noisy environment can be classified as an \emph{embarassing parallel} problem, since there is little to none communication among realizations. Problems of this kind take great advantage of GPGPU computing, since the solving algorithms can be designed to run directly on the GPU in such a way that communications are implemented via shared memory on the device (graphic card) and data transfer between the host (CPU) and the device and v.v.\ is limited to unavoidable input/output operations. 

In this paper, we have compared parallel algorithms for CTQWs evolution in a noisy environment based on the exact diagonalization of the Hamiltonian, the 4-th order Runge-Kutta integration method and the Taylor-series expansion of the evolution operator. Solutions that avoid the diagonalization of the Hamiltionian result in a lower computational cost and pave the way to highly parallelizable algorithms within the SIMT paradigm, thus leading to a straightforward implementation directly on the GPU. We have then benchmarked 4 NVIDIA GPUs and 3 quad-core Intel CPUs for a 2-particle system over a lattice of increasing dimensions and have shown that the GPU speedup with respect to the OPENMP parallelization fluctuates from 8x to more than 20x, depending on the frequency of post-processing. Thus, GPU-accelerated codes allow the design of simulations involving many particles or large lattices, with the only limit of the memory available on the device. 

The paper is organized as follows: In Sect.\ \ref{ALGO} we discuss and derive efficient algorithms for the dynamics of CTQWs; in Sect.\ \ref{IMPL} we provide the main details on their implementation and in Sect.\ \ref{RESULTS} we compare the performances of the algorithms on different CPUs and GPUs. Sect.\ \ref{CONC} closes the paper with some concluding remarks.

\section{Algorithms for quantum walks in a noisy environment}\label{ALGO}

Let us consider a $q$-dimensional regular lattice hosting $m$ quantum particles, and let $N_i$ and $2k_i$ 
be the numbers of mesh elements (nodes) and of neighbors to be considered along the $i$-th 
direction. The system in hand is described by an $N^m \times N^m$ matrix, storing the elements of the 
Hamiltonian $\mathcal{H}$ and by a $N^m$ vector for the wave-function $\Psi$, where $N=\prod_i N_i$ 
is the total number of mesh nodes. When $k=\sum_i 2k_i \ll N$, the Hamiltonian $\mathcal{H}$ is largely 
sparse with a maximum filling factor $\left(m k + 1\right) / N^m$. 

Since we are interested in quantum walks in a noisy environment, transitions from node $\alpha$ and to node $\beta$ are ruled by $k+1$ deterministic ($c_{\alpha\beta}$) and stochastic ($\xi_{\alpha\beta}$) parameters. The stochastic terms $\xi_{\alpha\beta}$ switch between multiple values at random times during the simulation (switching times) in order to describe (generally time-dependent) fluctuations 
induced by lattice imperfections and/or external sources of noise. 
Thus, the elements of the Hamiltonian $\mathcal{H}_{\alpha\beta}$ read 
\begin{equation}\label{Hij}
\mathcal{H}_{\alpha\beta}=\left\{ 
\begin{array}{ll}
c_{\alpha\beta} + \xi_{\alpha\beta} & \beta=\alpha \mbox{ or } \beta \mbox{ connected to } \alpha \\
0 & \mbox{otherwise}
\end{array}.
\right.
\end{equation}
The terms $\mathcal{H}_{\alpha\alpha}$ quantify the on-site energies of the walkers, whereas the terms $\mathcal{H}_{\alpha\beta}$  with $\alpha\neq \beta$ describe the tunneling amplitudes between neighboring sites (we assume that tunneling occurs only between neighboring nodes).

The time evolution of the system is provided by the Schr\"odinger equation
\begin{equation}\label {Schroedinger}
i \hbar \frac{\dd \ket{\Psi}}{\dd t}=\hat{\mathcal{H}}\ket{\Psi},
\end{equation}
where $\hbar$ is the reduced Planck constant; the knowledge of $\ket{\Psi(t)}$ at each time step yields the $N^m \times N^m$ density matrix $\boldsymbol{\rho}(t)=\ket{\Psi(t)} \bra{\Psi(t)}$, which is used to evaluate the average over realizations $\left\langle\boldsymbol{\rho}(t)\right\rangle$ and eventually further post-processed to calculate any desired observable quantity.

Consequently to the introduction of random terms, in order to avoid overweighting of outliers and produce a 
reliable ensemble average it is required to run a sufficiently large number $R$ of simulations (a.k.a. realizations, usually $R\ge1000$), and then averaging the density matrix.  In order to speed-up the calculation, and significantly cut the execution time, realizations can be run in parallel, as they are independent from each other. However, in the parallel execution memory usage rapidly increases because at least an Hamiltonian matrix $\mathcal{H}_i$ and a wave-function $\Psi_i$ must be stored for each realization $i$. As a matter of fact, memory occupancy may become quickly an issue when the grid size and/or the number of particles increase. 

\subsection{Diagonalization of the Hamiltonian}\label{DIAG}
If we suppose that the Hamiltonians $\mathcal{H}_i$ do not change significantly within the time-step $\delta t$, eq.\ \eqref{Schroedinger} can be solved in the quasi-static approximation. The \emph{exact} time evolution of a QW is provided by the well-known eigenproblem
\begin{equation}\label{eigenproblem}
(\hat{\mathcal{H}_i}-\mathcal{E}_i)\ket{\Psi_i}=0,
\end{equation}
that yields the eigenvalues $\boldsymbol{\varepsilon}_{ij}$ and the eigenvectors $\mathbf{w}_{ij}$ 
of the $i$-th Hamiltonian. The evolution of the wave function is then given by 
\begin{equation}\label{evolPsi}
\ket{\Psi_i(t+\delta t)} = \sum_j \exp\left(-\frac{i}{\hbar} \varepsilon_{ij} \delta t\right) \ket{\mathbf{w}_{ij}}\braket{\mathbf{w}_{ij}} {\Psi_i(t)}.
\end{equation}
The pseudocode for the parallel implementation is given in Algorithm \ref{algEigen}.
\begin{algorithm}[ht]
\caption{Pseudocode for solving the CTQW dynamics via diagonalization of the Hamiltonian matrix}
\begin{algorithmic}[1]
\State Initialize Hamiltonians $\mathcal{H}_i$
\State Initialize switching times
\While {time $t<t_{\rm{max}}$}
\ForAll {realizations}
\Comment {Begin Parallel Section}
\State Diagonalize $\mathcal{H}_i \rightarrow \{\boldsymbol{\varepsilon}_{ij}, \mathbf{w}_{ij}\}$ 
\State $\ket{\Psi_i(t+\delta t)} \leftarrow \sum_j e^{-\frac{i}{\hbar} \varepsilon_{ij} \delta t} \ket{\mathbf{w}_{ij}}\braket{\mathbf{w}_{ij}} {\Psi_i(t)}$ 
\State Update switching times
\State $\mathcal{H}_i\leftarrow\mathcal{H}_i(t+\delta t)$
\EndFor
\Comment {End Parallel Section, $\sim O(N^{3m})$}
\State $t \leftarrow t+\delta t$
\If {postprocessing}
\State $\langle \boldsymbol{\rho}(t)\rangle \leftarrow \frac1R \sum_i \ket{\Psi_i(t)}\bra{\Psi_i(t)}$ \Comment{$\sim O(N^{2m})$}
\State Post-process $\langle\boldsymbol{\rho}(t)\rangle$
\EndIf
\EndWhile 
\end{algorithmic}
\label{algEigen}
\end{algorithm}
\par
It is worth noticing that a) this algorithm requires a large number of computationally intensive events of the order $\sim O(N^{3m})$  and b) it is necessary to store $N^m$ eigenvectors of $N^m$ components per realization, which is exactly the same memory space that the dense Hamiltonian matrix would occupy. As a matter of fact, this issue may jeopardize the efficiency of the code, even in the case of a parallel implementation.

\subsection{Integration of ordinary differential equations}\label{RK} 

Going back  to the general solution of eq.\ \eqref{Schroedinger}, we may directly tackle the time-dependent Schr\"odinger equation as a set of ordinary differential equations for the vector $\ket{\Psi_i}$ and solve it by means of standard integration techniques that dispose of the calculation of the eigenstates. A widely-used integration scheme is represented by the 4th-order Runge-Kutta method.

In this case, there is no need of allocating a memory space as large as a dense Hamiltonian would require.  The Hamiltonian topology, i.e., how nodes are connected to each other, is known \emph{a-priori} from the definition of the mesh, and holds true for all of the realizations. In principle, up to $mk+1$ non-null elements are present in each row of the Hamiltonian. As a consequence, each of the $N^m \times N^m$ Hamiltonians $\mathcal{H}_i$ can be stored as a $N^m \times (mk+1)$ reduced matrix $\tilde{\mathcal{H}}_i$. A common  $N^m \times (mk+1)$ topology matrix holding the indexes of non-null elements also adds. Since transitions from node $\alpha$ to node $\beta$ and v.v.\ share the same rate, the symmetry of $\tilde{\mathcal{H}}_i$ allows for further memory savings down to $N^m \times (mk/2+1)$ elements.
These relationships hold true for a regular lattice; in the case of a general graph, where each site is connected to a variable number of other nodes, the approach is still applicable with the only difference that the number $k$ of non-null elements per row in the topology matrix is replaced by the number of connections.

The 4th-order Runge-Kutta procedure lets the wave-functions $\ket{\Psi_i}$ evolve by means of the linear  combination of 4 intermediate states $\ket{K_i^{(j)}}, j=1\ldots 4$. The evaluation of any component belonging to the $j$-th intermediate state requires only the knowledge of the wave-function at the current time step, the reduced Hamiltonian and the $(j-1)$-th state at the indexes stored in the corresponding row of the topology matrix. Since nodes are topologically equivalent to each other, SIMD and SIMT paradigms apply, allowing for a second degree of parallelization over nodes. The parallelizations over realizations and over nodes can be collapsed into a larger loop ($RN^m$ steps), which may better balance the computational burden assigned to each computing unit. The pseudocode for the implementation of the 4th-order Runge-Kutta method is reported in Algorithm \ref{algRK4}. 
\begin{algorithm}[ht]
\caption{Pseudocode for solving the CTQW dynamics via integration of the Schr\"odinger equation 
using the 4th-order Runge-Kutta method}
\begin{algorithmic}[1]
\State Define Hamiltonian topology
\State Initialize reduced Hamiltonians $\tilde{\mathcal{H}}_i$
\State Initialize switching times
\While {time $t<t_{\rm{max}}$}
\ForAll {realizations}
\Comment {Begin Parallel/SIMT Section}
\For {$j=1 \to 4$}
\State $\Big(\ket{\Psi_i},\ket{K_i^{(j-1)}},\tilde{\mathcal{H}}_i,\Big) \rightarrow \ket{K_i^{(j)}}$
\EndFor
\State $\ket{\Psi_i(t+dt)} \leftarrow \sum_{j=1}^4 \mu_j\ket{K_i^{(j)}}$ 
\State Check norm of $\ket{\Psi_i(t+dt)}$
\State Update switching times
\State $\tilde{\mathcal{H}}_i\leftarrow\tilde{\mathcal{H}}_i(t+\delta t)$ 
\EndFor
\Comment {End Parallel/SIMT Section, $\sim O(RN^m)$}
\State $t \leftarrow t+\delta t$
\If {postprocessing}
\State $\langle \boldsymbol{\rho}(t)\rangle \leftarrow \frac1R \sum_i \ket{\Psi_i(t)}\bra{\Psi_i(t)}$ \Comment{$\sim O(R N^{2m})$}
\State Post-process $\langle\boldsymbol{\rho}(t)\rangle$
\EndIf
\EndWhile 
\end{algorithmic}
\label{algRK4}
\end{algorithm}

The scheme in Algorithm \ref{algRK4} requires a single loop of sums and products; the algorithmic complexity of time evolution is thus reduced to the order $\sim O(RN^m)$, with a large speedup compared to the case discussed in Sect.\ \ref{DIAG}. The most computationally intensive routine is now represented by the calculation of the average density-matrix, order $\sim O(R N^{2m})$, whose number of calls may vary depending on the desired precision of the output.

The 4-th order Runge-Kutta method does not conserve the norm, and intermediate checks and corrective actions are required to avoid unphysical outcomes. It may also happen that the norm of $\ket{\Psi_i(t)}$ strongly deviates from its theoretical value within a single time step. In order to fix this issue two strategies may be devised. In the first one, higher-order Runge-Kutta methods can similarly be implemented to reach a better accuracy within the same time step, but memory allocation would grow since a larger number of intermediate states are required. On the other hand, one could reduce the time step in such a way that the cumulative error does not drive the simulation far away from its correct path. The immediate shortcoming is the increase of the running time inversely proportional to the step reduction; nonetheless, this solution becomes mandatory if it is not possible nor convenient to increase the memory allocation.


\subsection{Series expansion of the evolution operator}\label{SERIES}

Algorithm \ref{algRK4} may be modified in order to make the memory allocation independent of the required precision and slightly reduced with respect to the Runge-Kutta integration method. Upon introducing the evolution operator $\hat{\mathcal{U}}_i(\delta t)$, such that 
$$\ket{\Psi_i(t+\delta t)}= \hat{\mathcal{U}}_i(\delta t) \ket{\Psi_i(t)}\,,$$ 
we may rewrite Eq.\ \eqref{Schroedinger}  in terms of $\hat{\mathcal{U}}_i(\delta t)$ instead of $\ket{\Psi_i(t)}$, i.e.
\begin{equation}
i \hbar \frac{\dd \,\hat{\mathcal{U}}_i(\delta t)}{\dd t}=\hat{\mathcal{H}_i}\,\hat{\mathcal{U}}_i(\delta t)\,.
\end{equation}
The formal solution is given by $$\hat{\mathcal{U}}_i(\delta t)
=\exp\left(-\displaystyle{\frac{i}{\hbar}}\delta t\hat{\mathcal{H}_i}\right)\,.$$ 
Upon expanding $\hat{\mathcal{U}}_i(\delta t)$ in Taylor series we have
\begin{equation}
\begin{split}
\hat{\mathcal{U}}_i(\delta t)=&\mathbb{1}+\left(-\frac{i}{\hbar}\delta t\hat{\mathcal{H}_i}\right)
+\frac{1}{2}\left(-\frac{i}{\hbar}\delta t\hat{\mathcal{H}_i}\right)^2+ \dots +\\
&\qquad +\frac{1}{n!}\left(-\frac{i}{\hbar}\delta t\hat{\mathcal{H}_i}\right)^n +
o \left(\left(-\frac{i}{\hbar}\delta t\hat{\mathcal{H}_i}\right)^n\right);
\end{split}
\end{equation}
the wave-function can be recast as
\begin{equation}
\ket{\Psi_i(t+\delta t)}= \sum_{j=0}^n \ket{\Phi_i^{(j)}(t)}+o \left(\left(-\frac{i}{\hbar}\delta t\hat{\mathcal{H}_i}\right)^n\ket{\Psi_i(t)}\right),
\end{equation}
where $$\ket{\Phi_i^{(0)}(t)}=\ket{\Psi_i(t)}\,,$$ and $$\ket{\Phi_i^{(j)}(t)}=-\displaystyle{\frac{1}{j}\frac{i}{\hbar}\delta t\hat{\mathcal{H}_i}}\ket{\Phi_i^{(j-1)}(t)}\,.$$ 

The pseudocode for the evolution of the wave-functions by means of the expansion of the evolution operator in Taylor series is shown in Algorithm \ref{algSeries}.  In order to understand the similarities and the difference between the two methods,  let us remind that the coefficients $\mu_j$ in the Runge-Kutta expansion are, in general, determined by an educated fitting of a formal Taylor series expansion of the unknown functions in such a way that the truncation error is the same. Due to the exponential form of the evolution operator, there is also a perfect coincidence between the $n$-th order Taylor series expansion and the $n$-th order Runge-Kutta method~\cite{Butcher}. The advantage of the Taylor expansion is represented by the progressive updating of $\ket{\Psi_i(t)}$ with the help of the auxiliary vector $\ket{\Phi_i(t)}$, which is overwritten at each step of the expansion loop. Thus, the memory allocation of auxiliary variables does not depend any more on the precision of the calculation, without increasing the algorithmic complexity. Notice that at the same time, all the arguments discussed in Sect.\ \ref{RK}
about the heaviest routines (and the influence of the time step on the results)  still hold true.

\begin{algorithm}[ht]
\caption{Pseudocode for solving the CTQW dynamics via Taylor series expansion of the evolution operator}
\begin{algorithmic}[1]
\State Define Hamiltonian topology
\State Initialize reduced Hamiltonians $\tilde{\mathcal{H}}_i$
\State Initialize switching times
\While {time $t<t_{\rm{max}}$}
\ForAll {realizations}
\Comment {Begin Parallel/SIMT Section}
\State $\ket{\Phi_i^{(0)}(t)} \leftarrow \ket{\Psi_i(t)}$
\State $\ket{\Psi_i(t+\delta t)} \leftarrow \ket{\Psi_i(t)}$
\For {$j=1 \to n$}
\State $\ket{\Phi_i^{(j)}(t)} \leftarrow -\frac{1}{j}\frac{i}{\hbar}\delta t \hat{\mathcal{H}}_i \ket{\Phi_i^{(j-1)}(t)}$
\State $\ket{\Psi_i(t+\delta t)} \leftarrow \ket{\Psi_i(t+\delta t)} + \ket{\Phi_i^{(j)}(t)}$
\EndFor 
\State Check norm of $\ket{\Psi_i(t+dt)}$
\State Update switching times
\State $\tilde{\mathcal{H}}_i\leftarrow\tilde{\mathcal{H}}_i(t+\delta t)$ 
\EndFor
\Comment {End Parallel/SIMT Section, $\sim O(RN^m)$}
\State $t \leftarrow t+\delta t$
\If {postprocessing}
\State $\langle \boldsymbol{\rho}(t)\rangle \leftarrow \frac1R \sum_i \ket{\Psi_i(t)}\bra{\Psi_i(t)}$ \Comment{$\sim O(R N^{2m})$}
\State Post-process $\langle\boldsymbol{\rho}(t)\rangle$
\EndIf
\EndWhile 
\end{algorithmic}
\label{algSeries}
\end{algorithm}

\section{Implementation}\label{IMPL}

Algorithms \ref{algEigen}-\ref{algSeries} have been implemented to run on multicore shared-memory workstations and graphic accelerators, making use for linear algebra of the BLAS and LAPACK or the cuBLAS and CULA~\cite{Humphrey10} libraries on the host system and on the device, respectively. We have not tackled any advanced memory optimization: as it will be discussed in Sect.\ \ref{RESULTS}, benefits brought in by a highly-optimized code are not expected to further increase the performance gain significantly.

As far as Algorithm \ref{algEigen} is concerned, we envisage two workflows for parallel execution. On the one hand, would memory not be an issue, one can split realizations among non-communicating cores in such a way that, even though the single realization is serialized, a number of realizations are handled at the same time. On the other hand, it may be convenient to serialize realizations and decrease the single-realization running time by spreading the matrix diagonalizations and the matrix-matrix products on multiple cooperating cores. In principle, the latter solution can be pushed farther if a large number of computing cores are available to the programmer, as it is the case of GPUs. 

\begin{figure}
\centering
\includegraphics*[width=0.8\columnwidth]{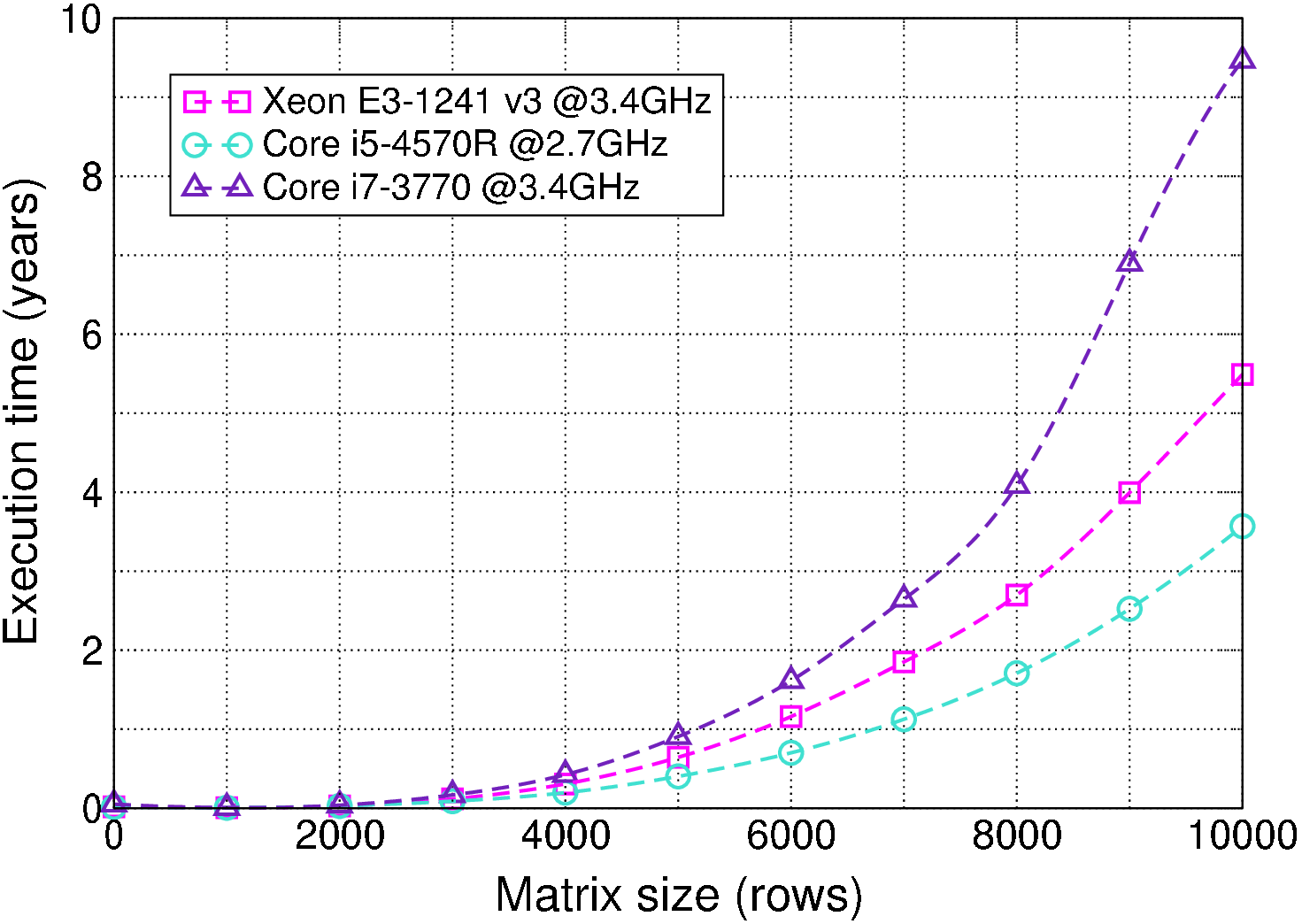}
\caption{Projected execution time required for diagonalizing $10^6$ times a symmetric matrix in single precision with the MKL 11.2 library. The tests have been performed on 4 CPU cores, as this configuration preliminarily proved to maximize the overall performance.}
\label{fig:TempoDiag}
\end{figure}

The execution times required by the diagonalization of symmetric matrices with single precision data (\texttt{ssyev} function of the Intel MKL 11.2 library) have preliminarily been measured for 3 Intel processors, then the outcomes have been projected over $10^6$ calls, which is the typical number of diagonalizations required for the problem in hand. As shown in {\figurename} \ref{fig:TempoDiag}, a simulation may last for years, which is a virtually infinite time for a computational physics problem. According to CULA white-papers \cite{Humphrey10}, the corresponding routine ported to GPUs may achieve a speedup ranging from 3x to 10x, a condition that still prevents any investigation from completing within an affordable time. 

Algorithms \ref{algRK4} and \ref{algSeries} have been implemented by means of 15 kernels directly on the GPU, then the corresponding OPENMP versions have been derived by replacing kernel invocations with loops. This approach allows for a direct execution time comparison since the number of floating-point operations is basically the same between host and device execution.

The two algorithms share the same 4-stage workflow (\emph{1.\ initialization; 2.\ wave-function evolution; 3.\ Hamiltonian update; 4.\ density-matrix calculation and post-processing}) and approximately 90\% of the code. Contrarily to Algorithm \ref{algEigen}, where the limiting factor is primarily represented by time, the limiting factor of Algorithms \ref{algRK4} and \ref{algSeries} is given by the memory required to store the (symmetric, complex) density-matrix $\langle \boldsymbol{\rho}(t) \rangle$ and the wave-functions $\ket{\Psi_i(t)}$. Top level, high-performance solutions for GPGPU computing like NVIDIA Tesla K80 offer up 24 GB of GPU-RAM, which cap the maximum size around 51000 rows (e.g., $q=2$, $m=2$, $N_1\cdot N_2=225$). 

\section{Performance evaluation}\label{RESULTS}

In order to evaluate the performance of Algorithms \ref{algRK4} and \ref{algSeries} we tested the case of 
unidimensional, 2-particle, nearest-neighbor CTQWs with periodic boundary conditions (i.e., $q=1$, $m=2$, $k=2$) and random noise on the tunneling energies. Simulations of 1500 time steps for $R=1000$ realizations, with different rates for post-processing (from 1 out of 1500 to 1 out of 10 time steps) have been run on the following hardware:

\begin{itemize}
\item {\bf Intel CPU}: Core i5-4570R @ 2.7 GHz and 8 GB RAM (4 cores), OS X 10.10.5
\item  {\bf Intel CPU}: Core i7-3770 @ 3.4 GHz and 4 GB RAM (4 cores), 64-bit Linux OS 
\item  {\bf Intel CPU}: Xeon E3-1241 v3 @ 3.5 GHz and 16 GB RAM (4 cores), 64-bit Linux OS 
\item {\bf NVIDIA GPU}: Tesla M2050 with 3 GB VRAM, ECC enabled, Compute capability 2.0, CUDA Toolkit 5.0
\item  {\bf NVIDIA GPU}: Tesla K40 with 12 GB VRAM, ECC enabled, Compute capability 3.5, CUDA Toolkit 7.5
\item  {\bf NVIDIA GPU}:  Tesla K80 with 24 GB VRAM, ECC enabled, Compute capability 3.7, CUDA Toolkit 7.5
\item  {\bf NVIDIA GPU}:  GeForce GTX980 with 4 GB VRAM, no ECC, Compute capability 5.2, CUDA Toolkit 6.5
\end{itemize}

The OPENMP source code has been compiled with the Intel C++ Compiler (ICC) version 15.0.3 for Linux and version 15.0.7 for OS X; the CUDA source code with the NVIDIA CUDA Compiler (NVCC), with no further optimizations other than those provided by default. Preliminary runs on GPUs proved that 256 threads per block maximize the efficiency.

\begin{figure}
\centering
\includegraphics*[width=0.8\columnwidth]{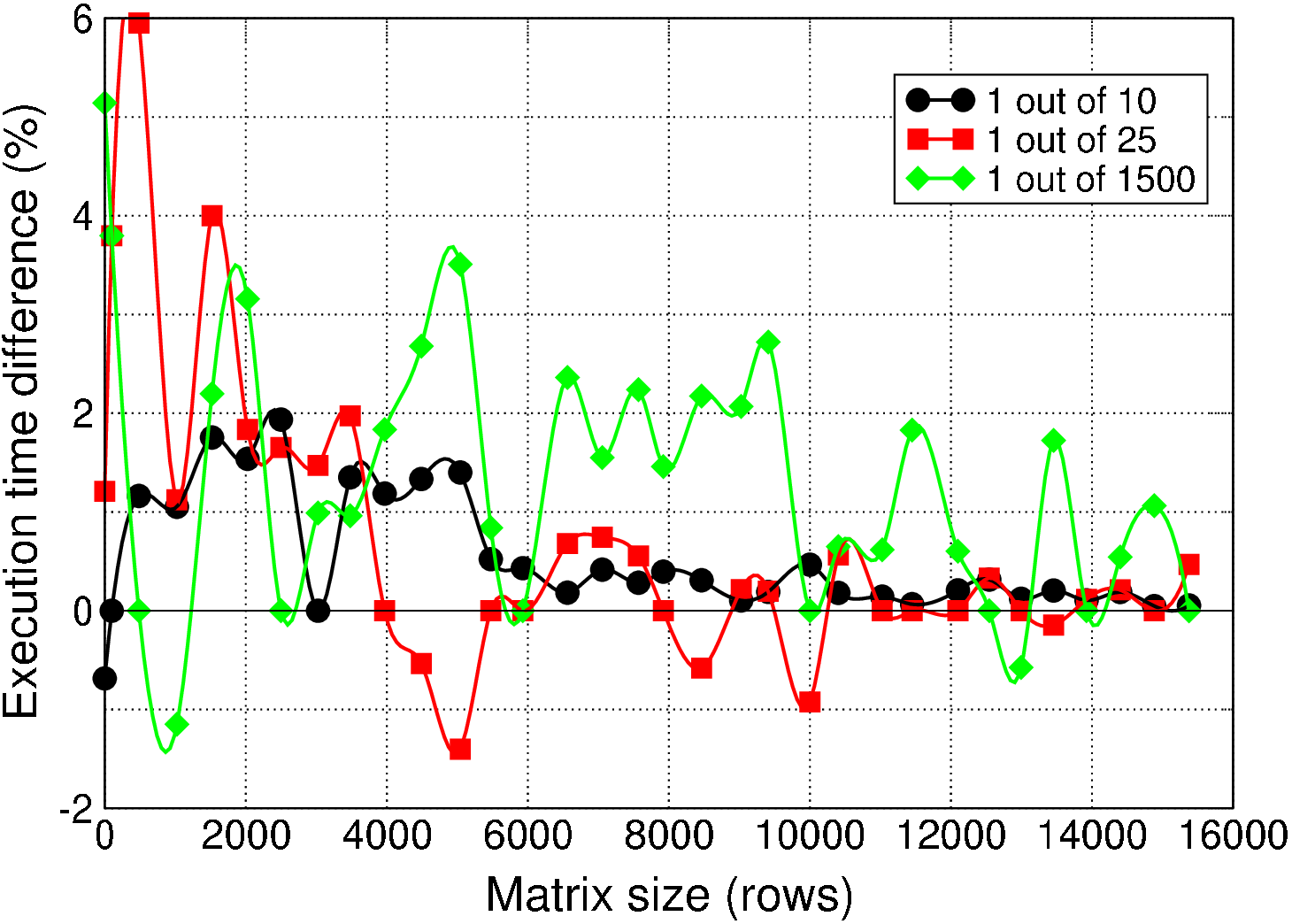}
\caption{Runtime difference between Algorithms \ref{algRK4} and \ref{algSeries} on a K40 board as a function of the post-processing rate. The 4-th order Runge Kutta method is on average 2 seconds faster, which becomes a negligible time as long as the size of the problem increases.}
\label{fig:RKvEXP}
\end{figure}

The execution times of the 4-th order Runge-Kutta and of the series expansion methods are basically the same. Depending on the hardware, very few seconds in favor of one algorithm or the other are reported;  differences become negligible as long as the size of the mesh increases (see {\figurename} \ref{fig:RKvEXP} for tests performed on the Tesla K40). Therefore, we proceed in the analysis only with Algorithm \ref{algSeries} and assume that the same conclusions hold true also for Algorithm \ref{algRK4}.

\begin{figure}
\centering
\subfigure[]{\includegraphics*[width=0.49\columnwidth]{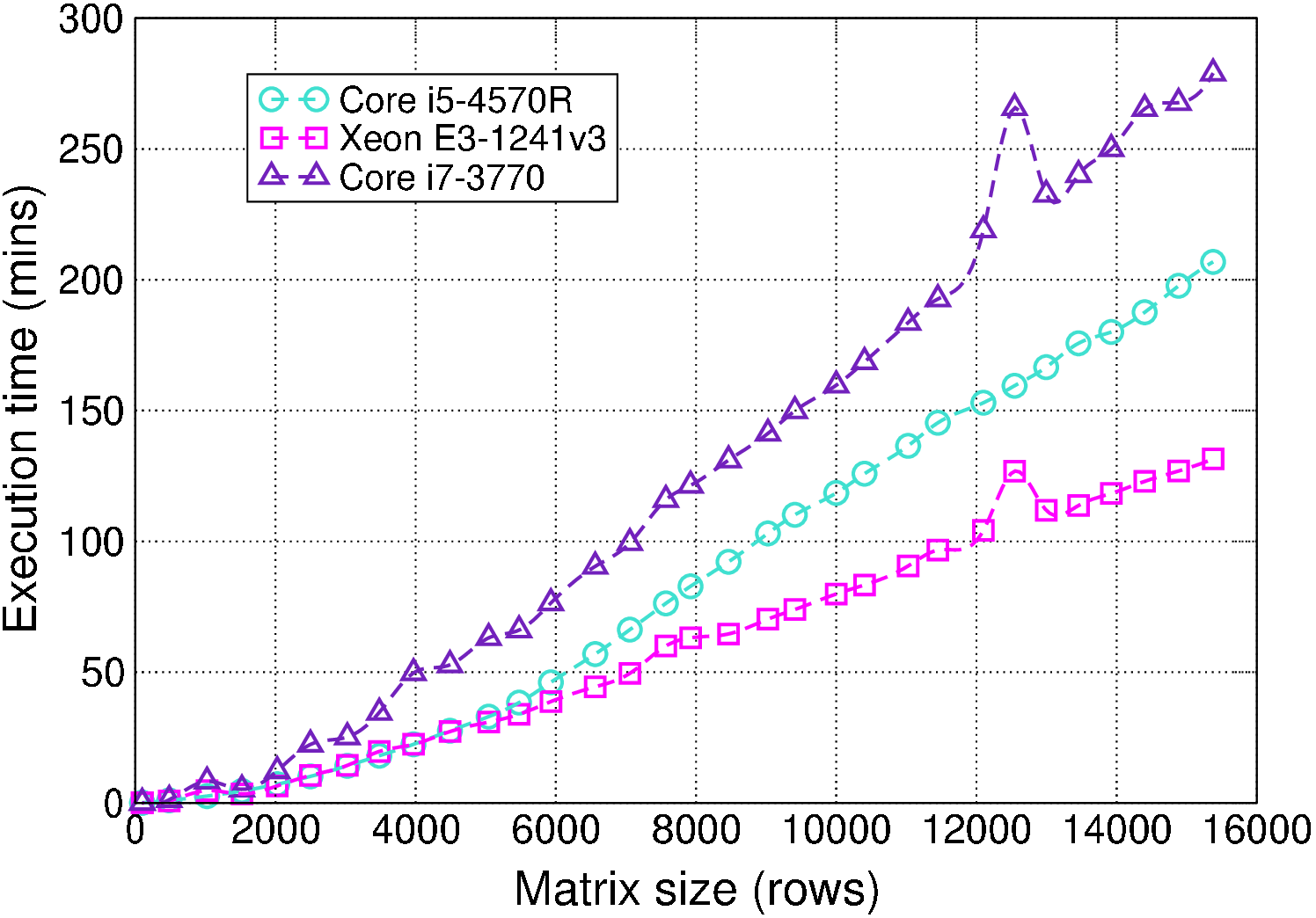}\label{CPU_10}}
\hfill
\subfigure[]{\includegraphics*[width=0.49\columnwidth]{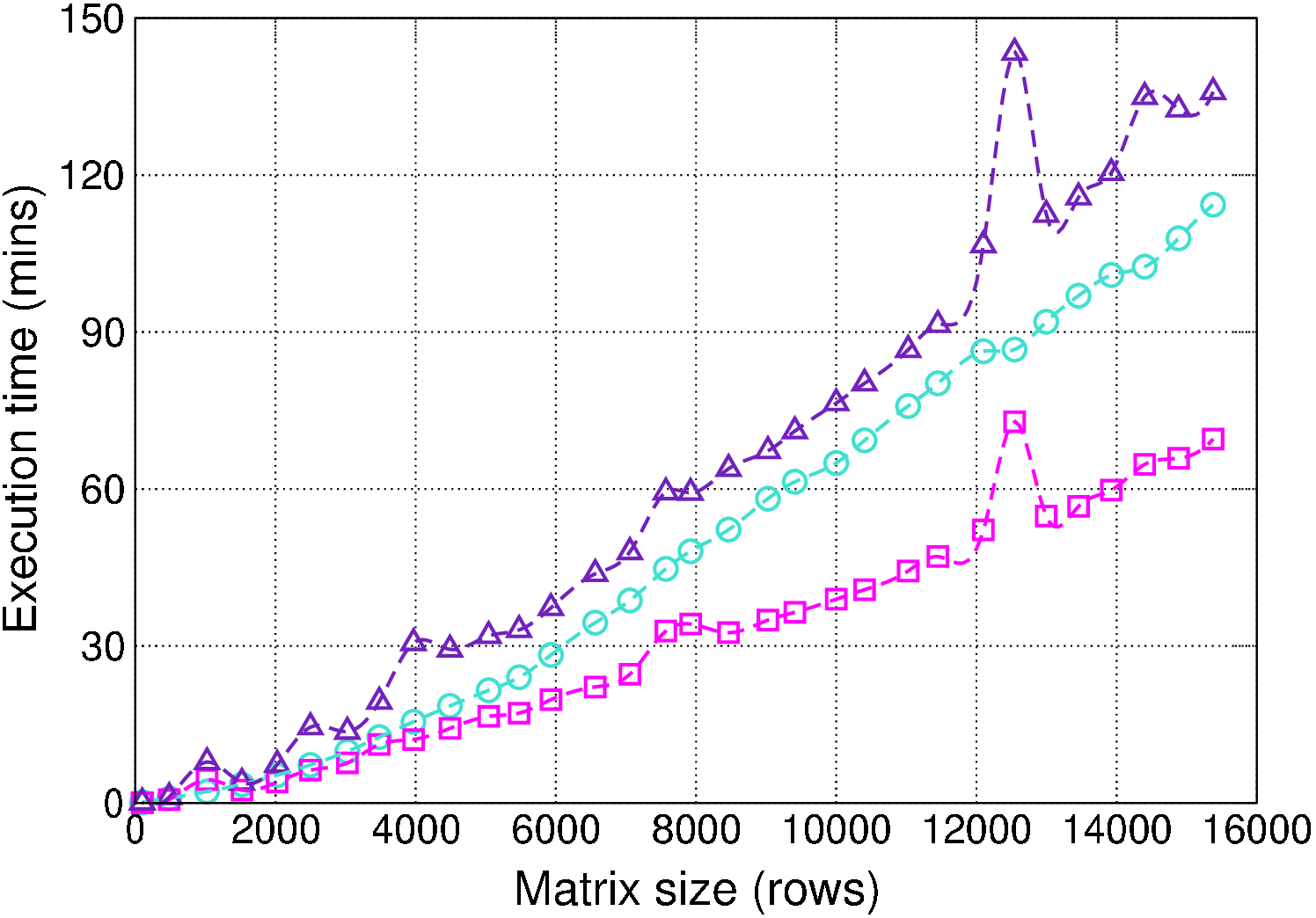}\label{CPU_25}}\\
\subfigure[]{\includegraphics*[width=0.49\columnwidth]{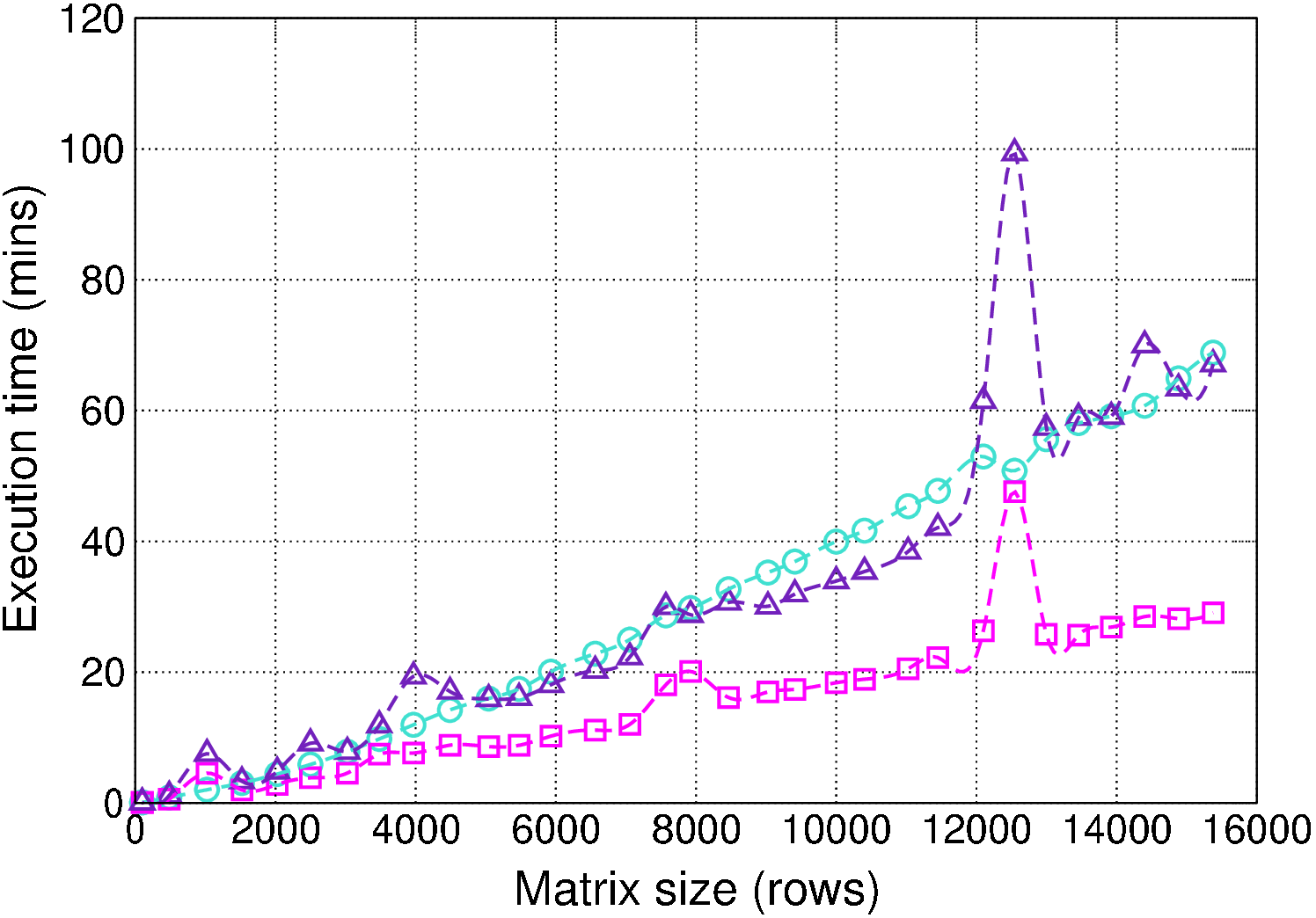}\label{CPU_100}}
\hfill
\subfigure[]{\includegraphics*[width=0.49\columnwidth]{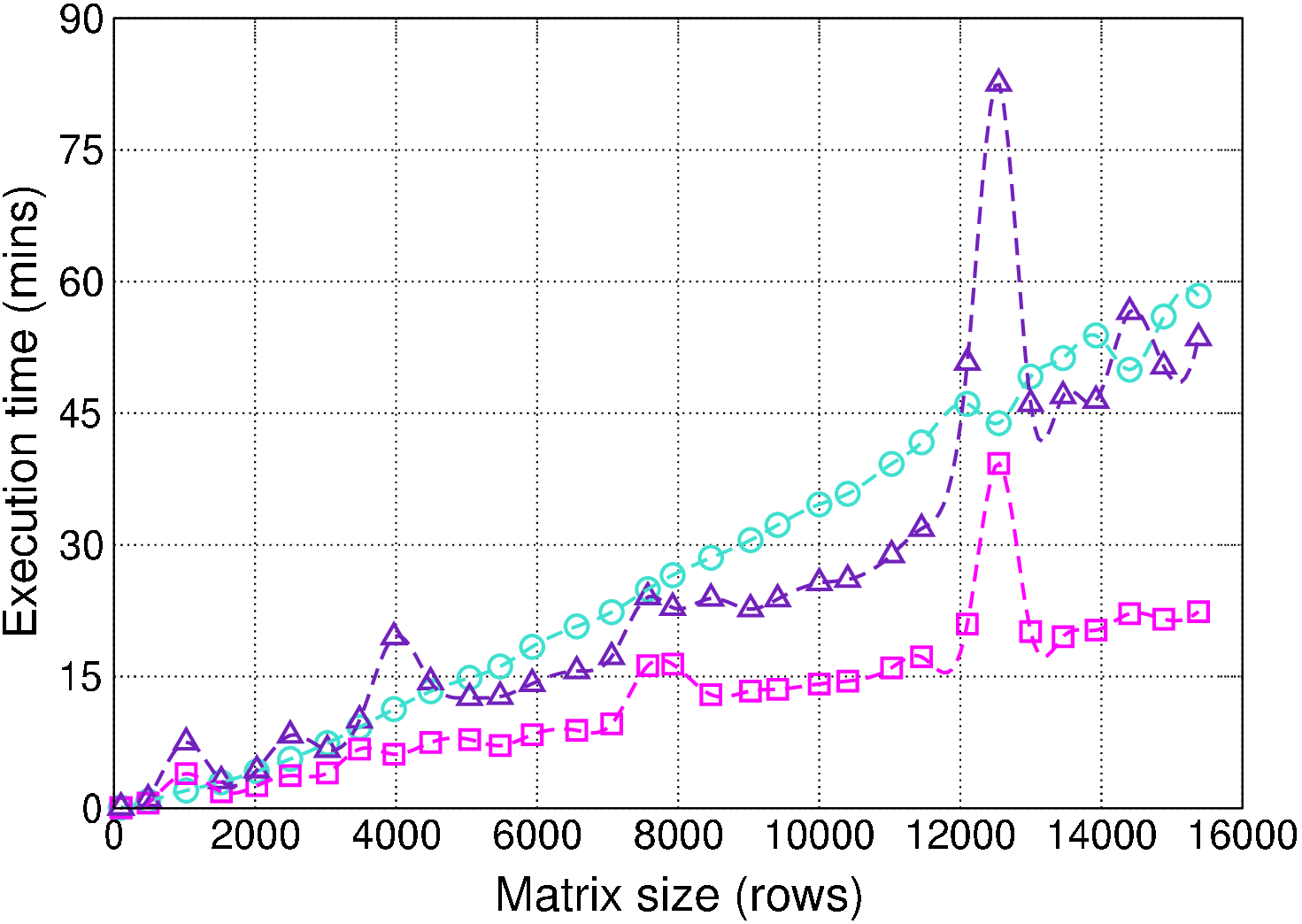}\label{CPU_250}}\\
\subfigure[]{\includegraphics*[width=0.49\columnwidth]{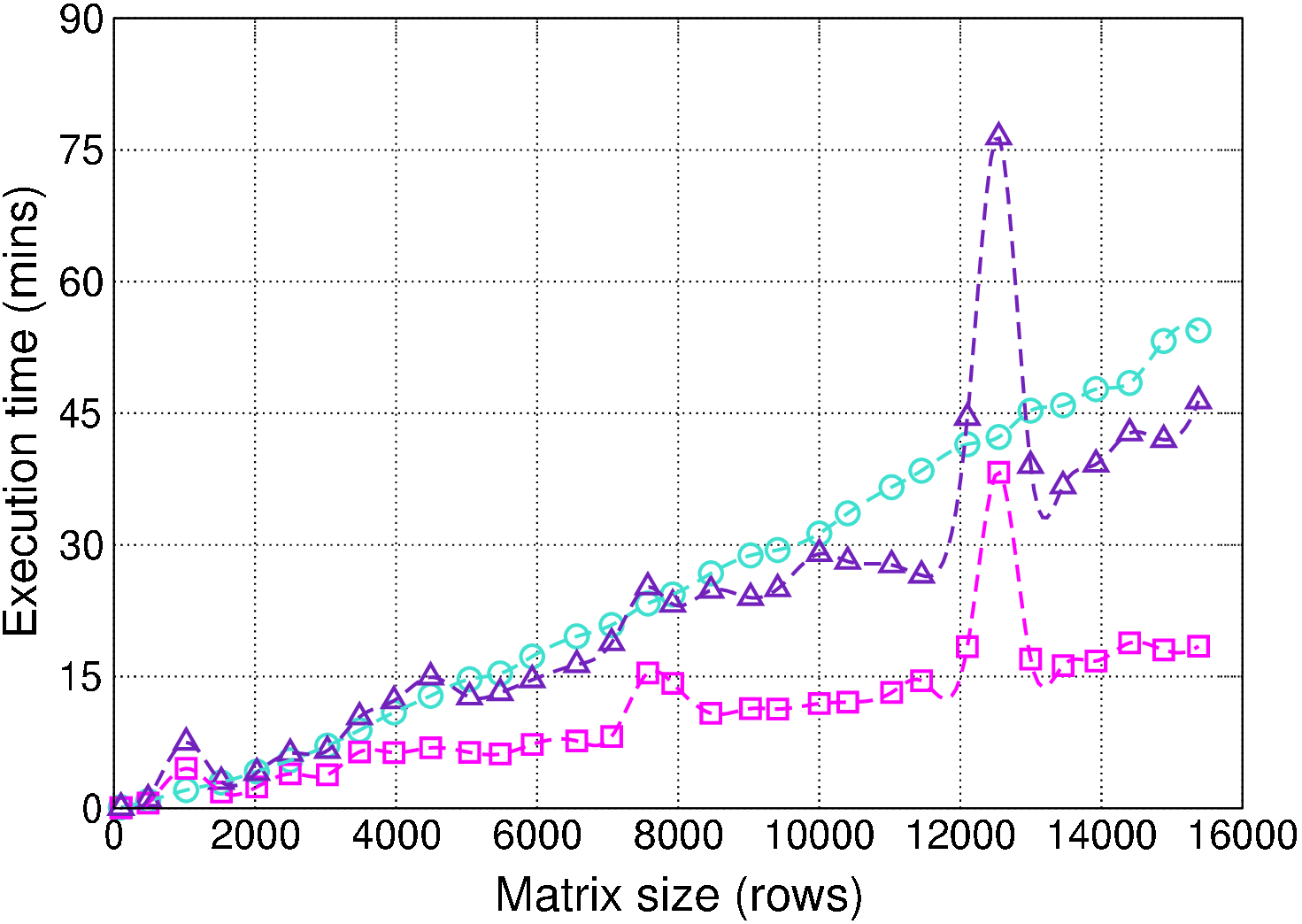}\label{CPU_1500}}
\hfill
\subfigure[]{\includegraphics*[width=0.49\columnwidth]{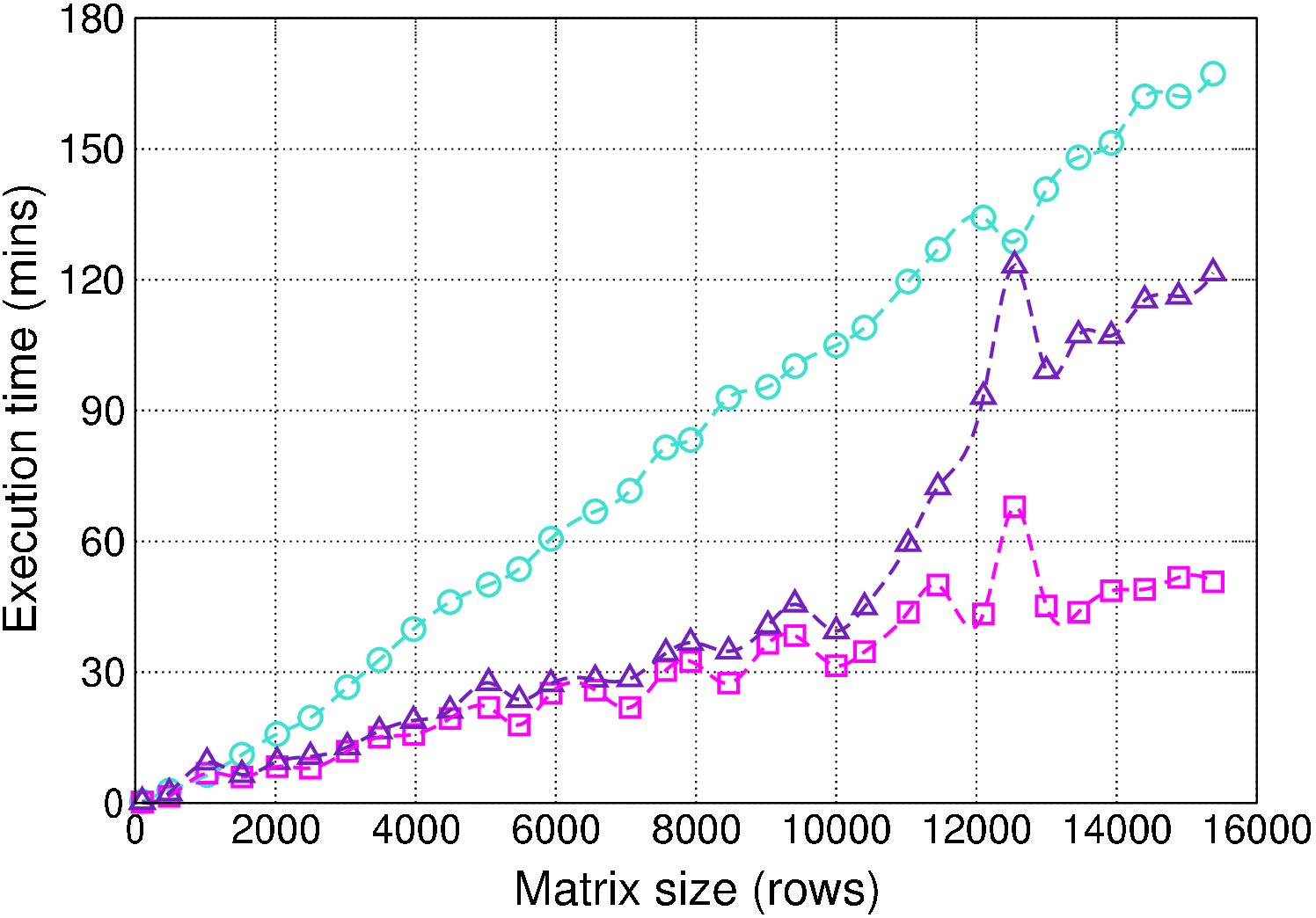}\label{CPU_ser}}
\caption{Execution time of the OPENMP code for different rates of post-processing. \ref{CPU_10}: 1/10 time steps; \ref{CPU_25}: 1/25 time steps; \ref{CPU_100}: 1/100 time steps; \ref{CPU_250}: 1/250 time steps; \ref{CPU_1500} and \ref{CPU_ser}: 1/1500 time steps. Panel \ref{CPU_ser} refers to the execution time of the single-core case.
}
\label{fig:CPU}
\end{figure}

{\figurename} \ref{fig:CPU} illustrates the execution time of the Series expansion algorithms as a function of the problem size for the three CPUs under test. A strong performance loss sets in around 12500 rows for the Core-i7 and the Xeon E3 processors, that, more generally, also share an evident qualitative correlation of the running times as a function of the mesh size. Since these evidences are not confirmed in the Core-i5 case, we attribute this underperformance to a compiler issue specific of version 15.0.3. 

\begin{figure}
\centering
\subfigure[]{\includegraphics*[width=0.49\columnwidth]{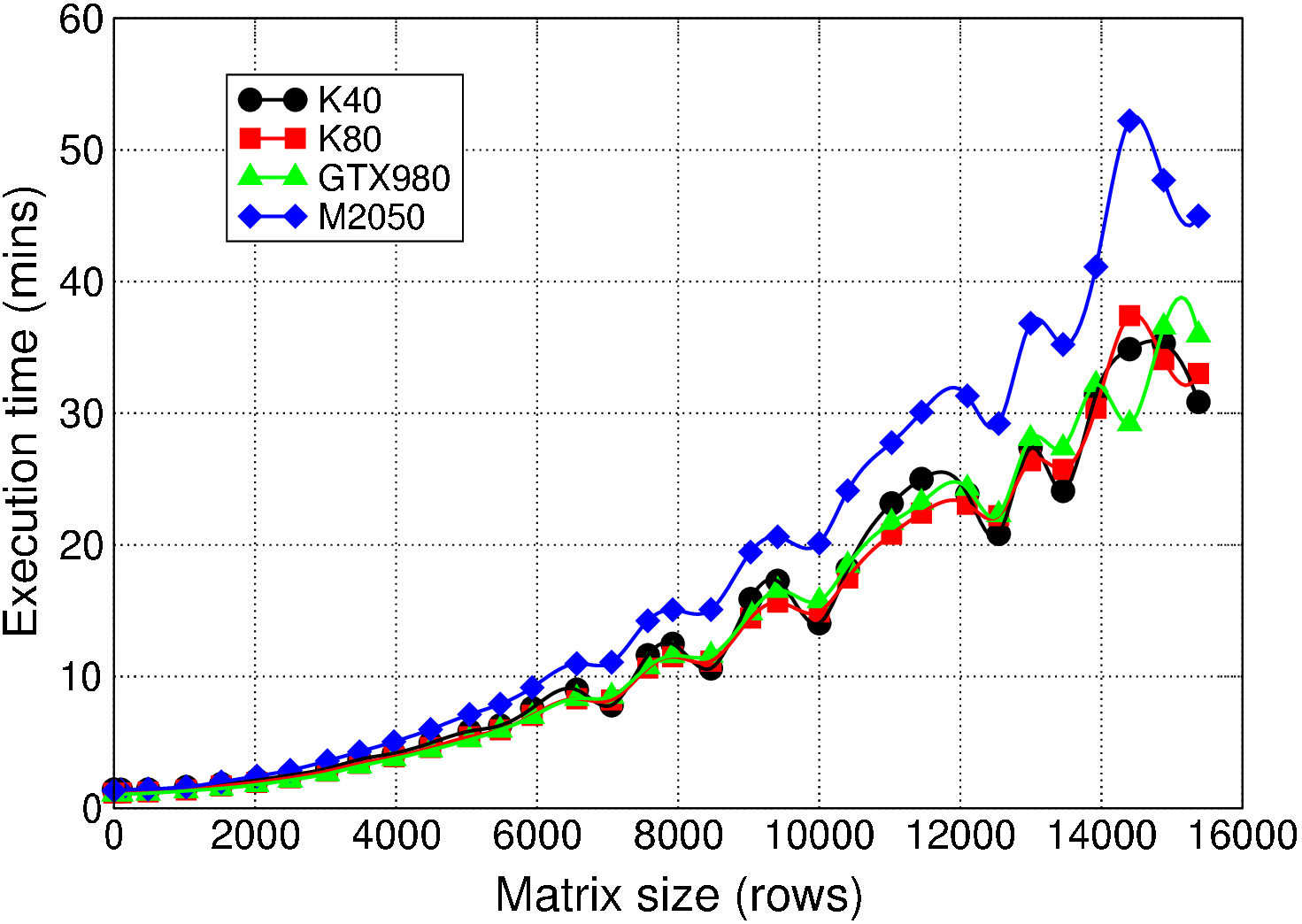}\label{GPU_10}}
\hfill
\subfigure[]{\includegraphics*[width=0.49\columnwidth]{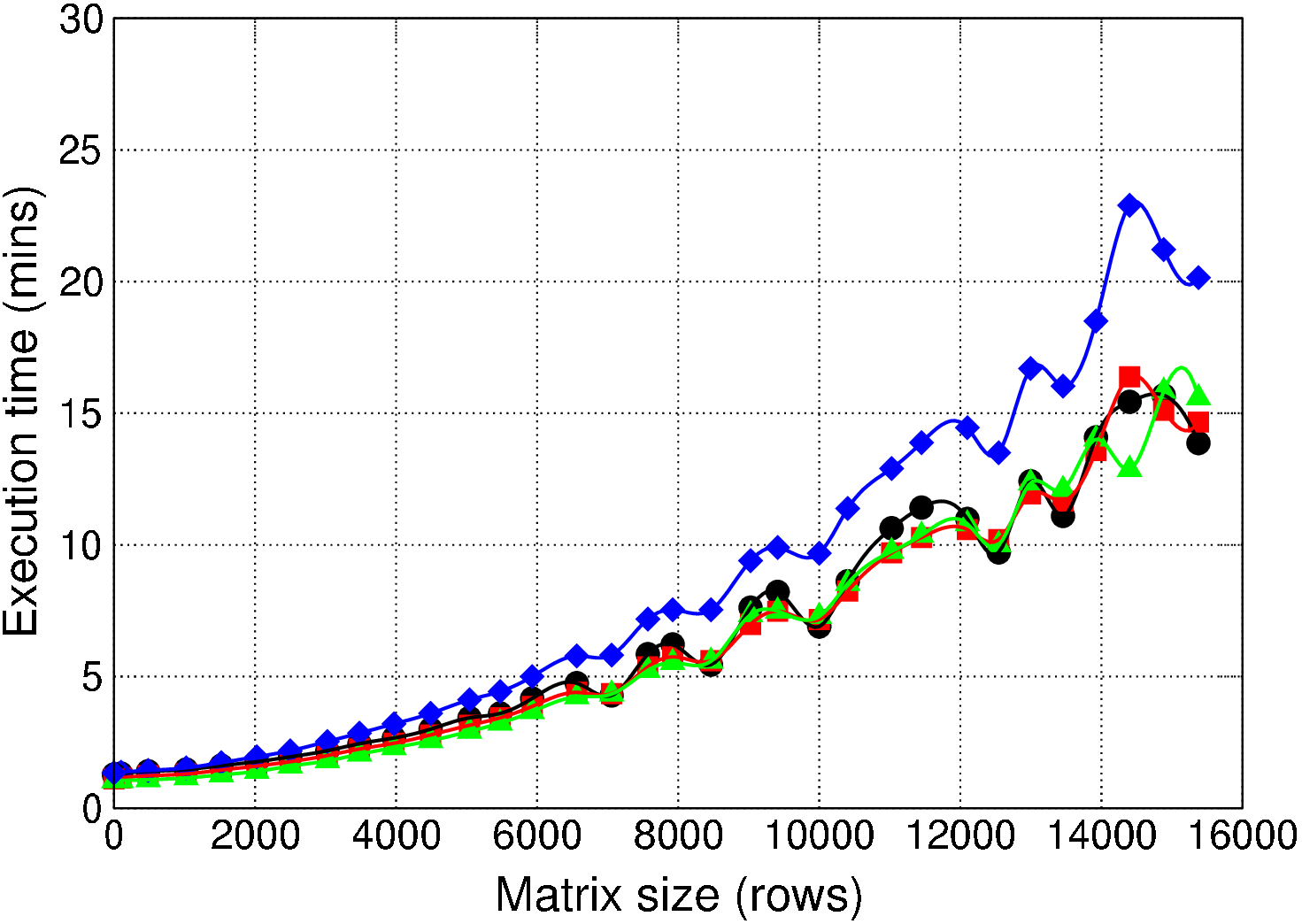}\label{GPU_25}}\\
\subfigure[]{\includegraphics*[width=0.49\columnwidth]{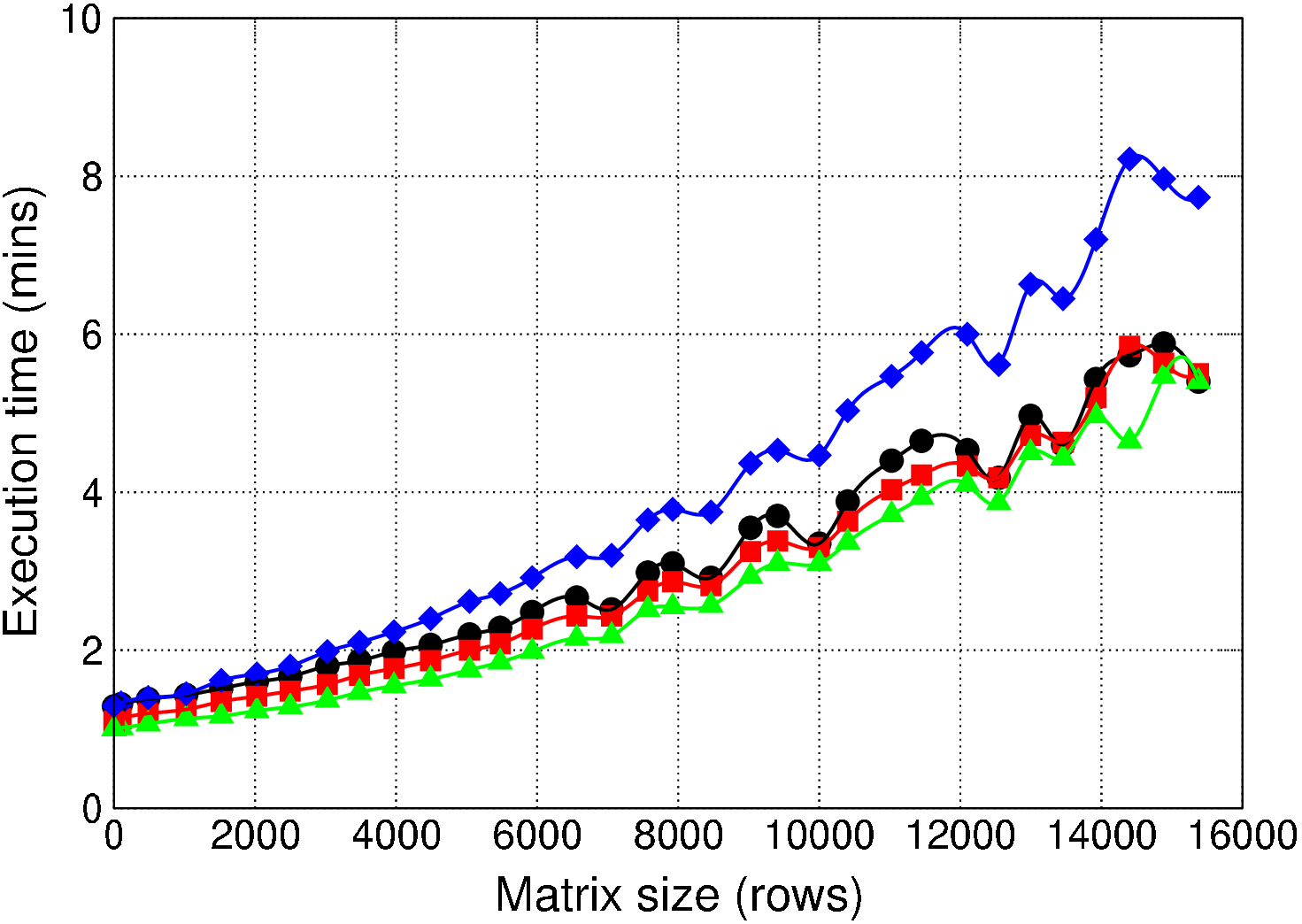}\label{GPU_100}}
\hfill
\subfigure[]{\includegraphics*[width=0.49\columnwidth]{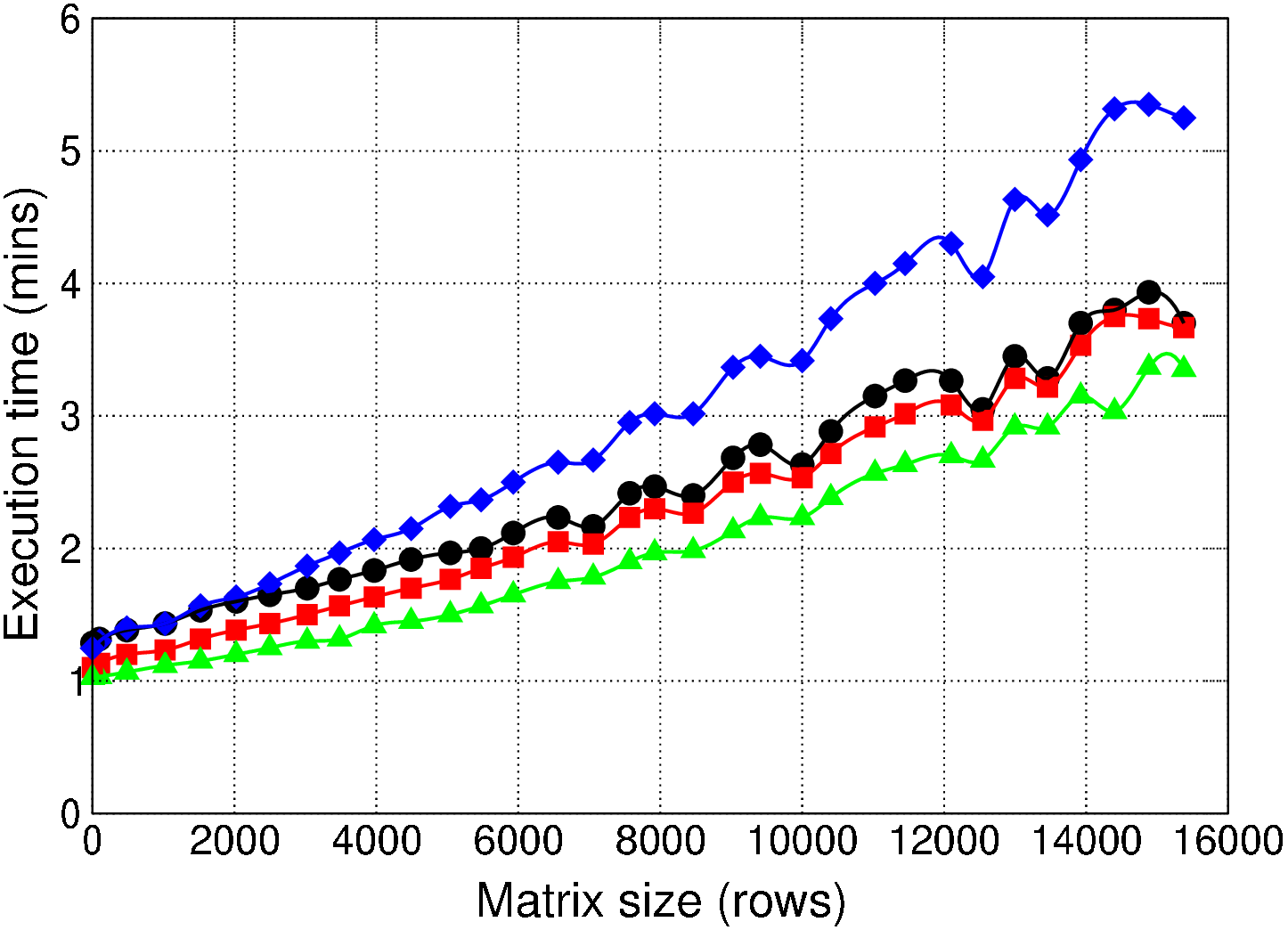}\label{GPU_250}}\\
\subfigure[]{\includegraphics*[width=0.49\columnwidth]{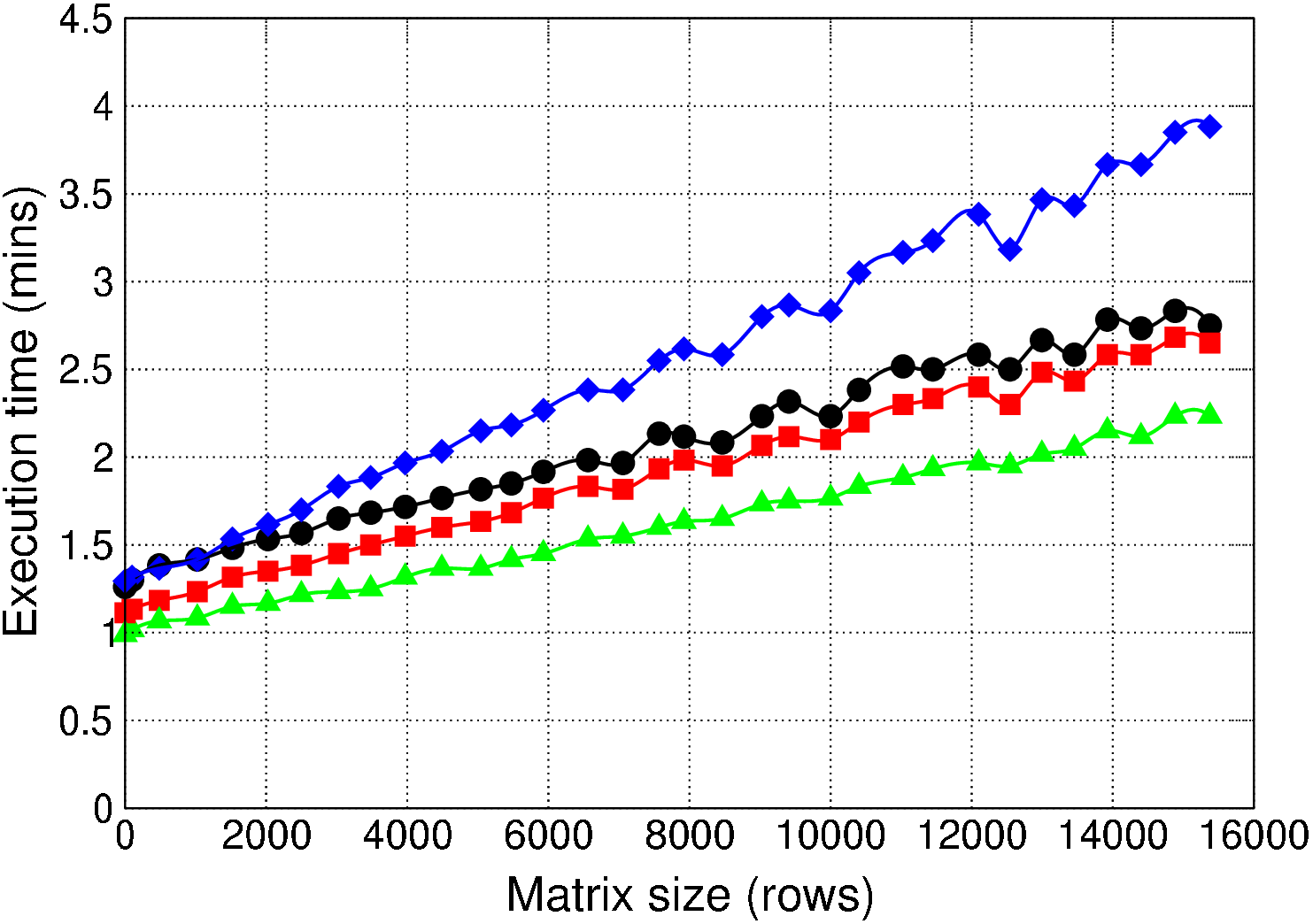}\label{GPU_1500}}
\caption{Execution time of the GPU code for different rates of post-processing. \ref{GPU_10}: 1/10 time steps; \ref{GPU_25}: 1/25 time steps; \ref{GPU_100}: 1/100 time steps; \ref{GPU_250}: 1/250 time steps; \ref{GPU_1500}: 1/1500 time steps.
}
\label{fig:GPU}
\end{figure}

The analogous execution time comparison for the codes running on the four GPUs is shown in {\figurename} \ref{fig:GPU}. All tests are completed in less than 1 hour, with running times very similar to each other. The only exception is represented by the Tesla M2050 board that underperforms its competitors, though retaining a substantial gain over any OPENMP execution. In general, the fastest runs are achieved with the GeForce GTX980 board thanks to a superior clock rate. Notice that GeForce boards are not certified for GPGPU computing due to lack of ECC memory, and uncontrolled bit-flips or erratic bits in the memory locations devoted to the storage of $\ket{\Psi_i(t)}$ can jeopardize the reliability of the outcomes of the simulations. Though uncommon, this aspect deserves care and double checks are mandatory in presence of odd results.

\begin{figure}
\centering
\subfigure[]{\includegraphics*[width=0.49\columnwidth]{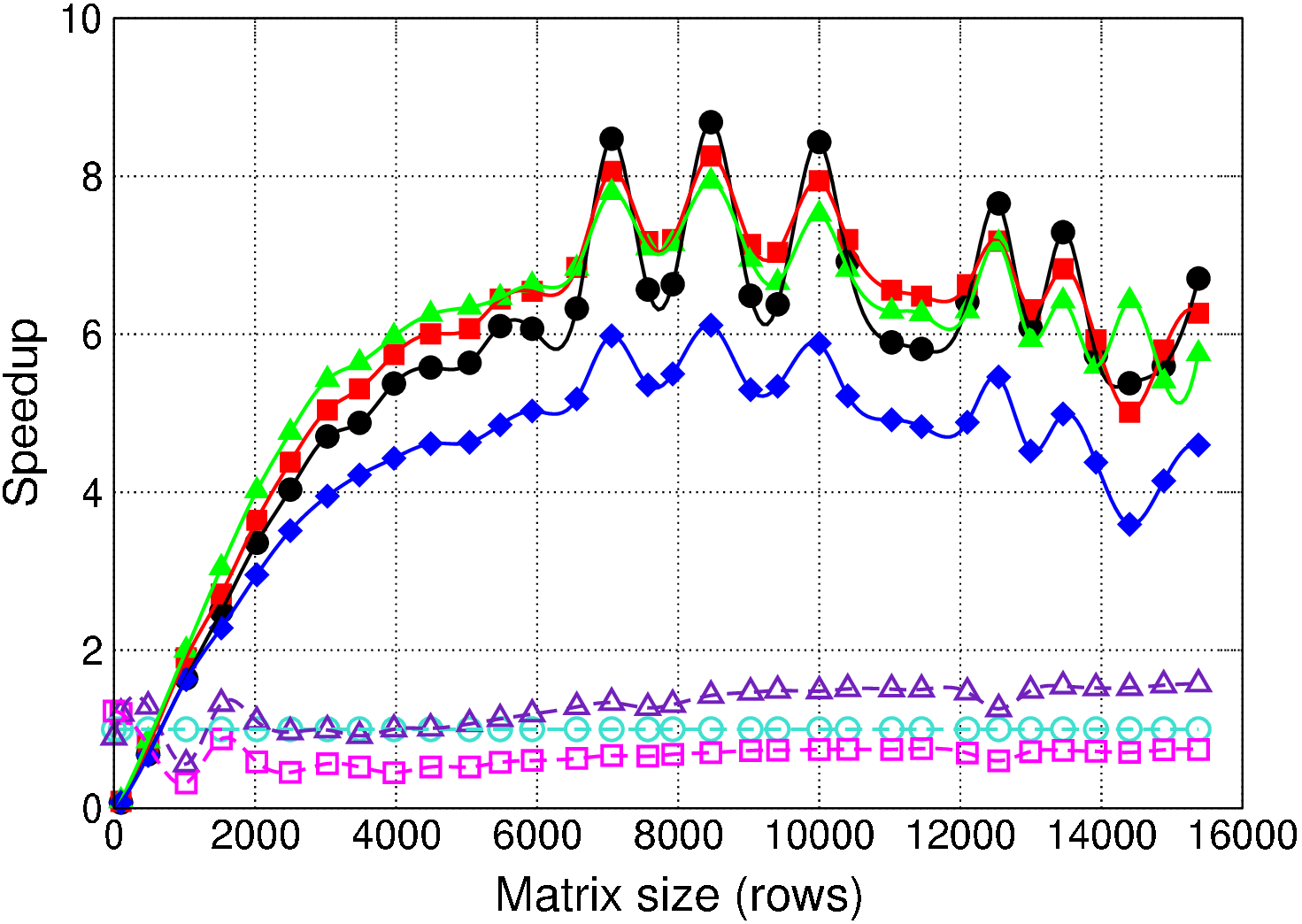}\label{SU_10}}
\hfill
\subfigure[]{\includegraphics*[width=0.49\columnwidth]{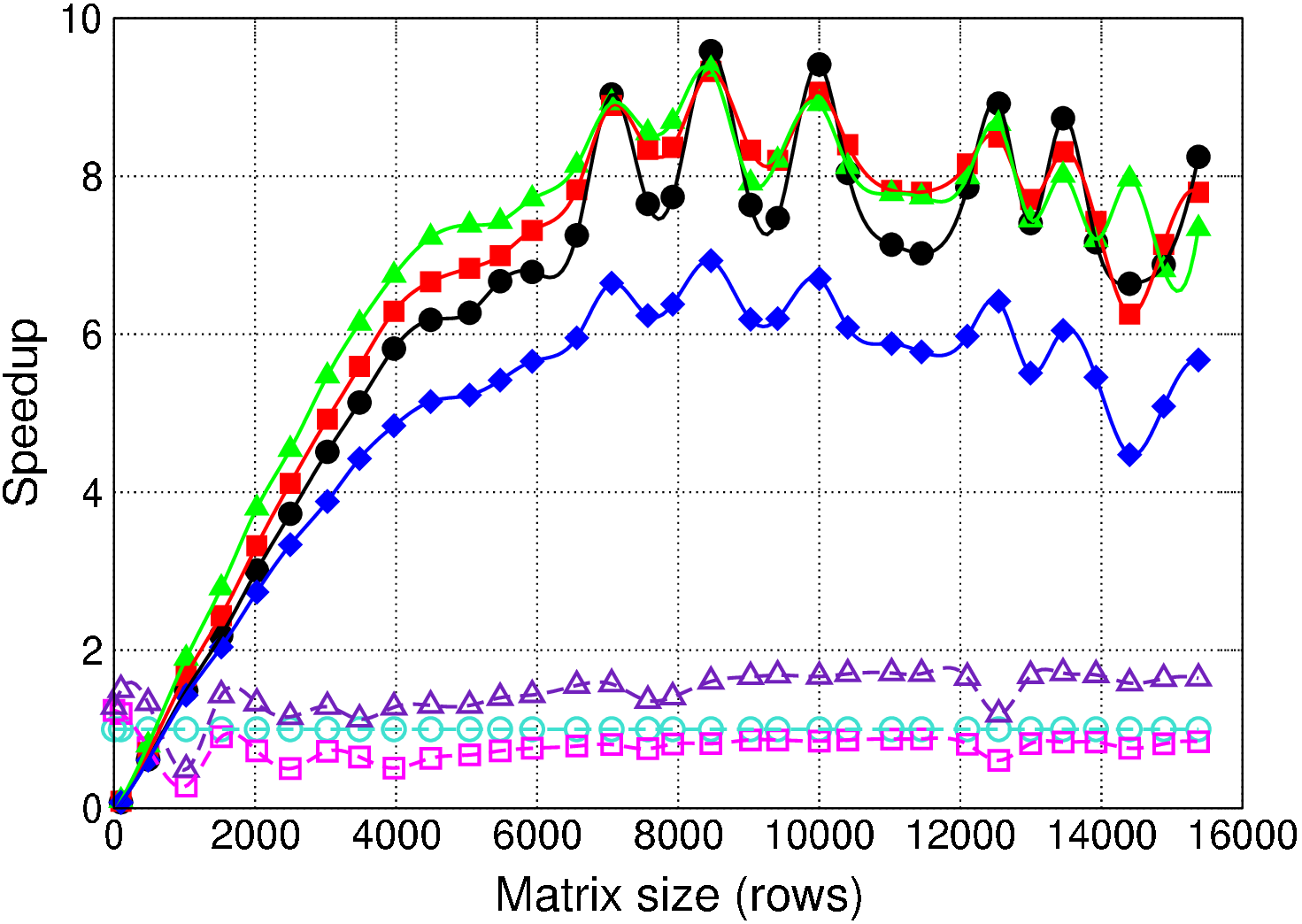}\label{SU_25}} \\
\subfigure[]{\includegraphics*[width=0.49\columnwidth]{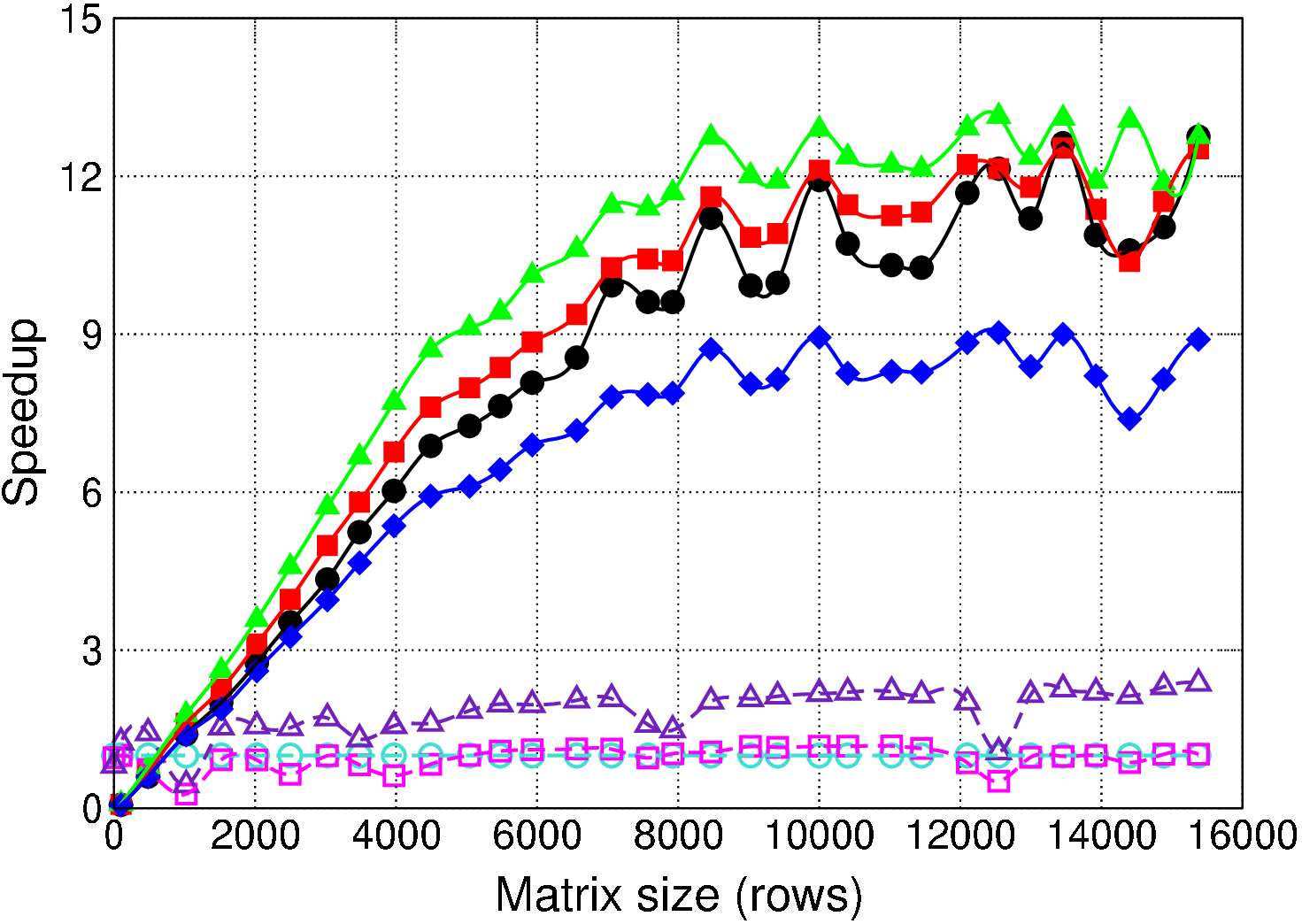}\label{SU_100}}
\hfill
\subfigure[]{\includegraphics*[width=0.49\columnwidth]{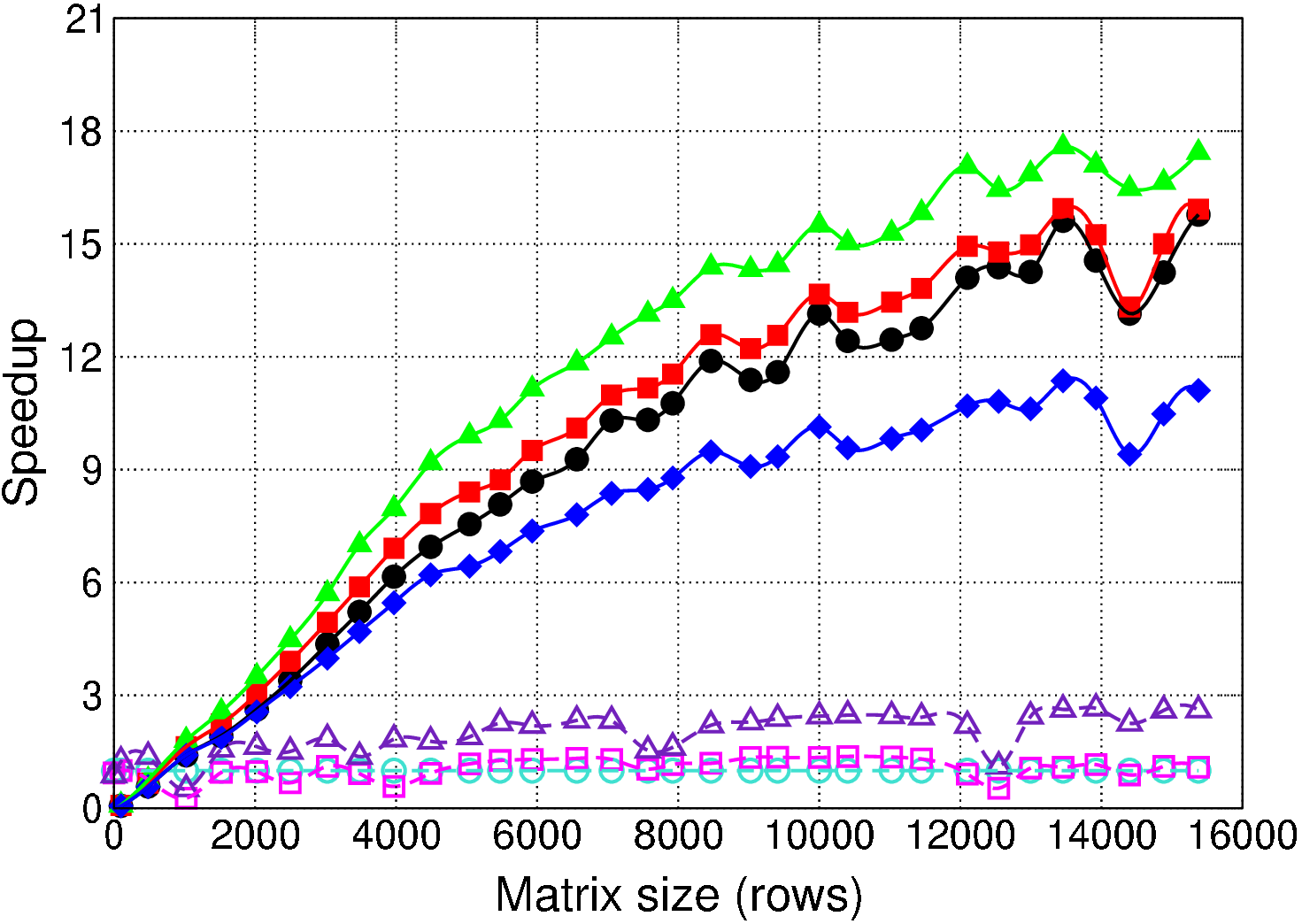}\label{SU_250}}\\
\subfigure[]{\includegraphics*[width=0.49\columnwidth]{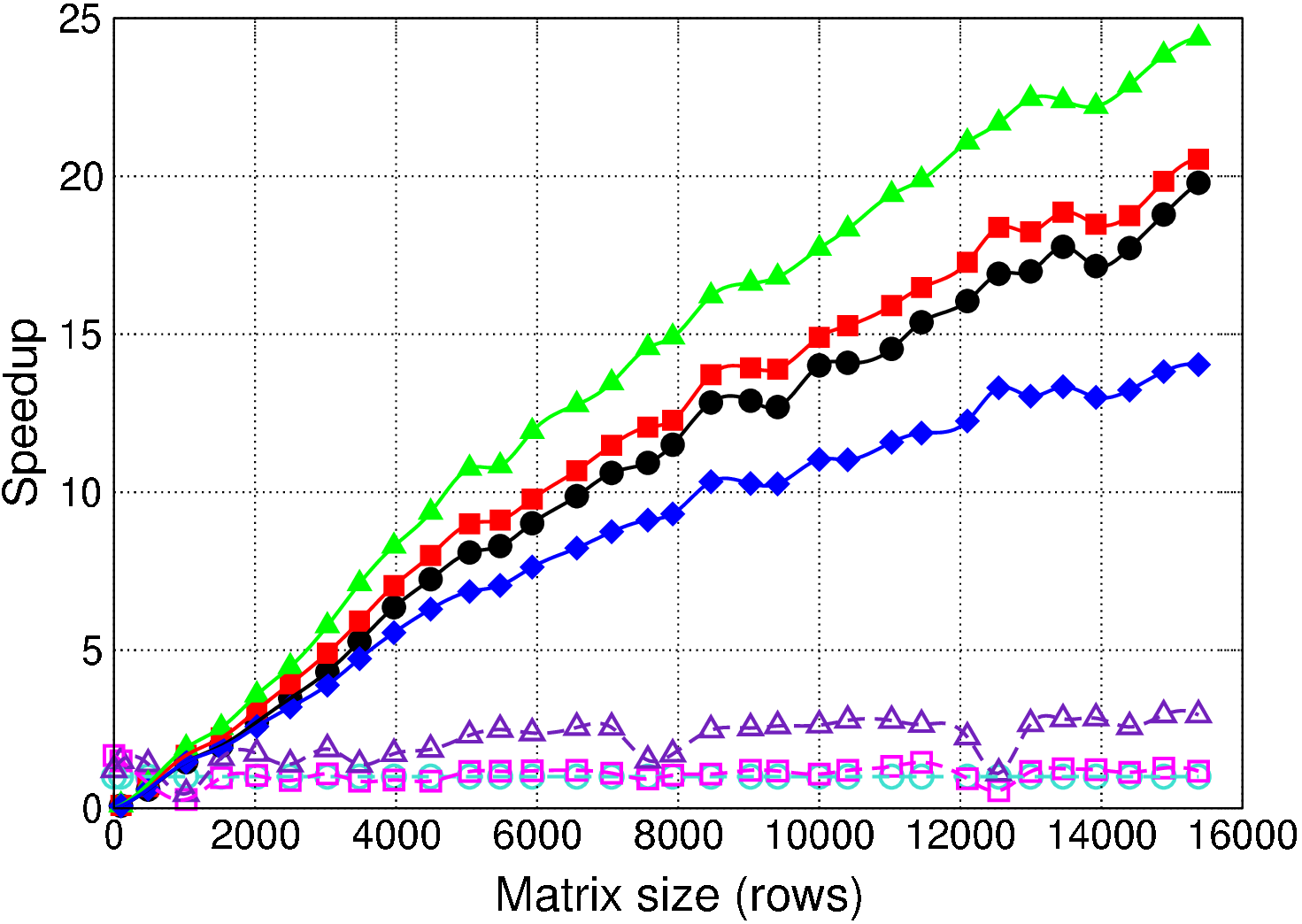}\label{SU_1500}}
\hfill
\subfigure[]{\includegraphics*[width=0.49\columnwidth]{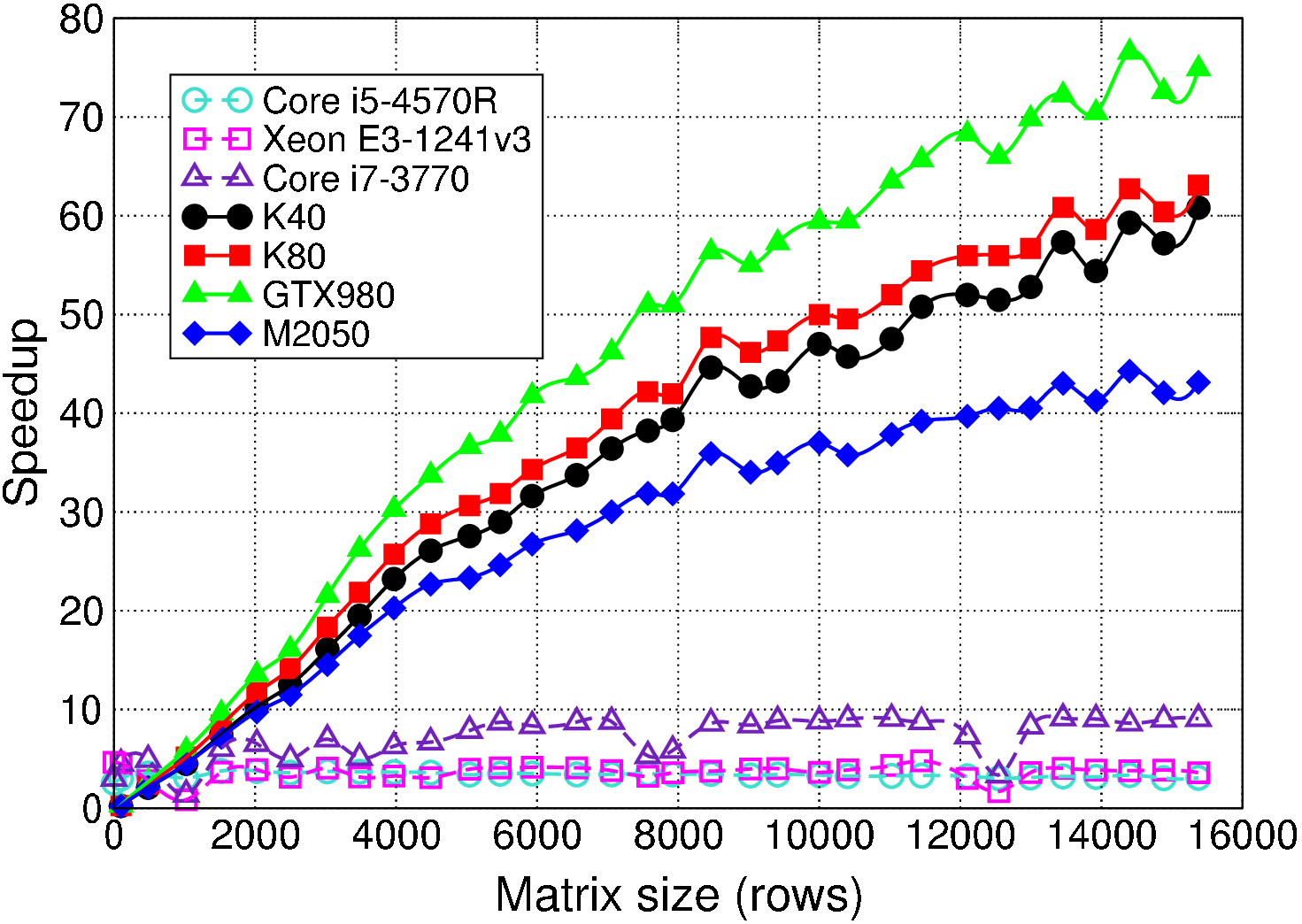}\label{SU_ser}}
\caption{Performance comparison for different rates of post-processing. \ref{SU_10}: 1/10 time steps; \ref{SU_25}: 1/25 time steps; \ref{SU_100}: 1/100 time steps; \ref{SU_250}: 1/250 time steps; \ref{SU_1500} and \ref{SU_ser}: 1/1500 time steps. In panel \ref{SU_ser} the comparison at post-processing rate 1/5000 time steps refers to the single-core execution.}
\label{fig:SU}
\end{figure}

In order to provide an overall comparative review of the performances, we chose the Core-i5 as the reference processor and we calculate the simulation speedup as
\begin{equation}
\mbox{Speedup}=\frac{\mbox{CPU or GPU-under-test execution time}}{\mbox{Core i5-4570R 4-thread execution time}}.
\end{equation}
Data are shown in {\figurename} \ref{fig:SU}. GPU implementation becomes convenient roughly about 1000 rows, when the workload starts to fill completely the computational power of the GPUs. Since this is a very common case for many-particle CTQWs (i.e., a mesh as small as $N\sim32$ for the two-particle case), GPU computing sounds a viable and efficient option to pursue in order to reduce the execution time down to the minute-to-few-hour range. It is important to stress that the simulation speedup strongly depends on the post-processing rate. For an output generation as frequent as 1 out of 10 time steps (panel a) a gain about 5x-7x is obtained; the gain rises up to 8x-9x for an average post-processing rate of 1 out of 25 time steps (panel b) and up to a 10x-13x for a moderate output generation around 1 out of 100 time steps (panel c). Panels d and e refer instead to cases where the calculation of the density matrix is progressively reduced down to a single time per simulation. In other words, this is the the speedup achievable for the pure evolution of the wave-functions, which settles in the 20x range and more. By comparing data reported in panels \ref{CPU_1500} and \ref{CPU_ser} of {\figurename} \ref{fig:CPU}, the OPENMP parallelization introduces a further 2.5-3x gain with respect to the single-core execution, this boosting up the speedup at a minimum gain around 60x for the pure evolution of the wave functions, as shown in {\figurename} \ref{SU_ser}, and around 15x when a high post-processing frequency is required.

\begin{figure}
\centering
\subfigure[]{\includegraphics*[width=0.49\columnwidth]{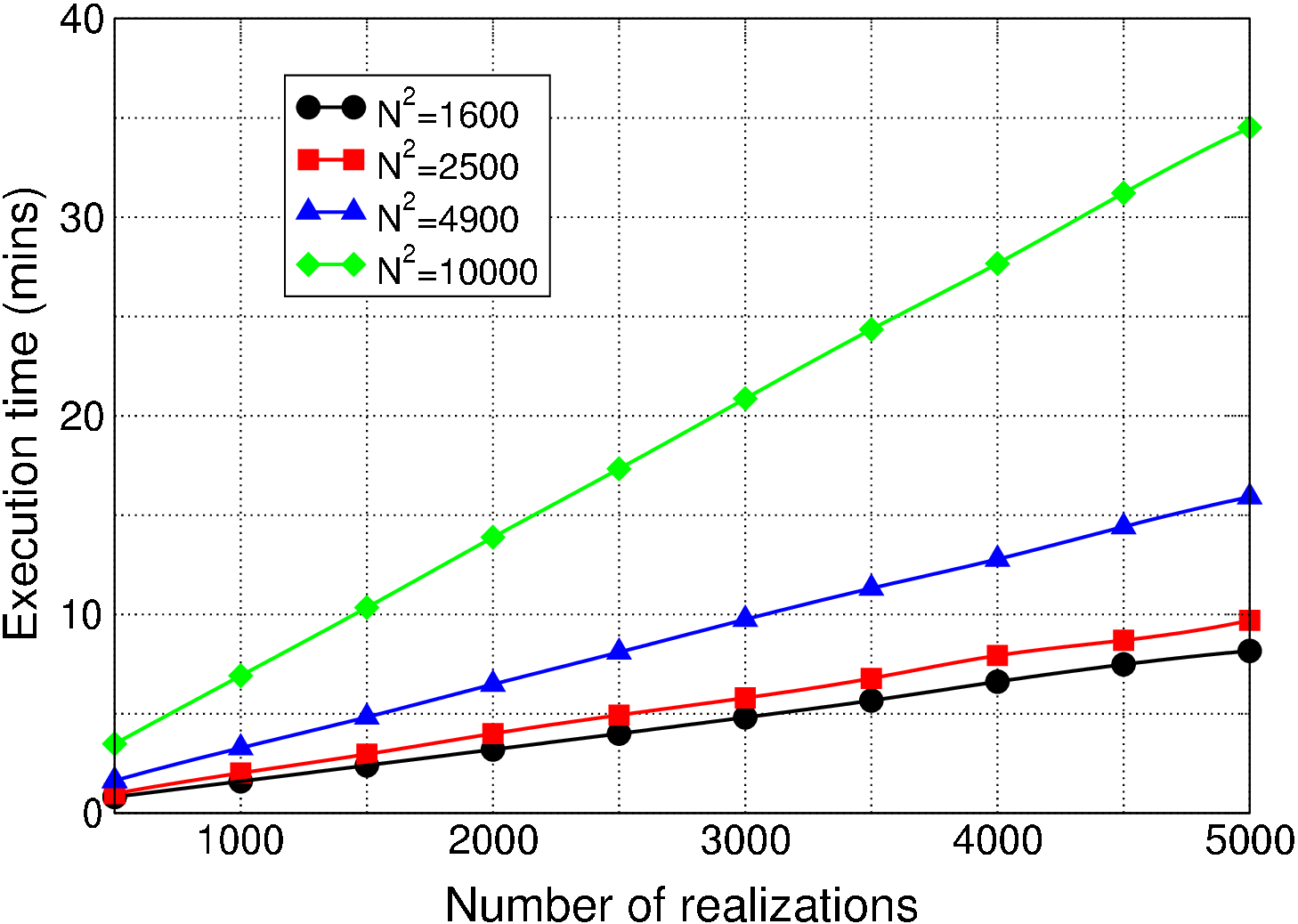}\label{GPU}}
\hfill
\subfigure[]{\includegraphics*[width=0.49\columnwidth]{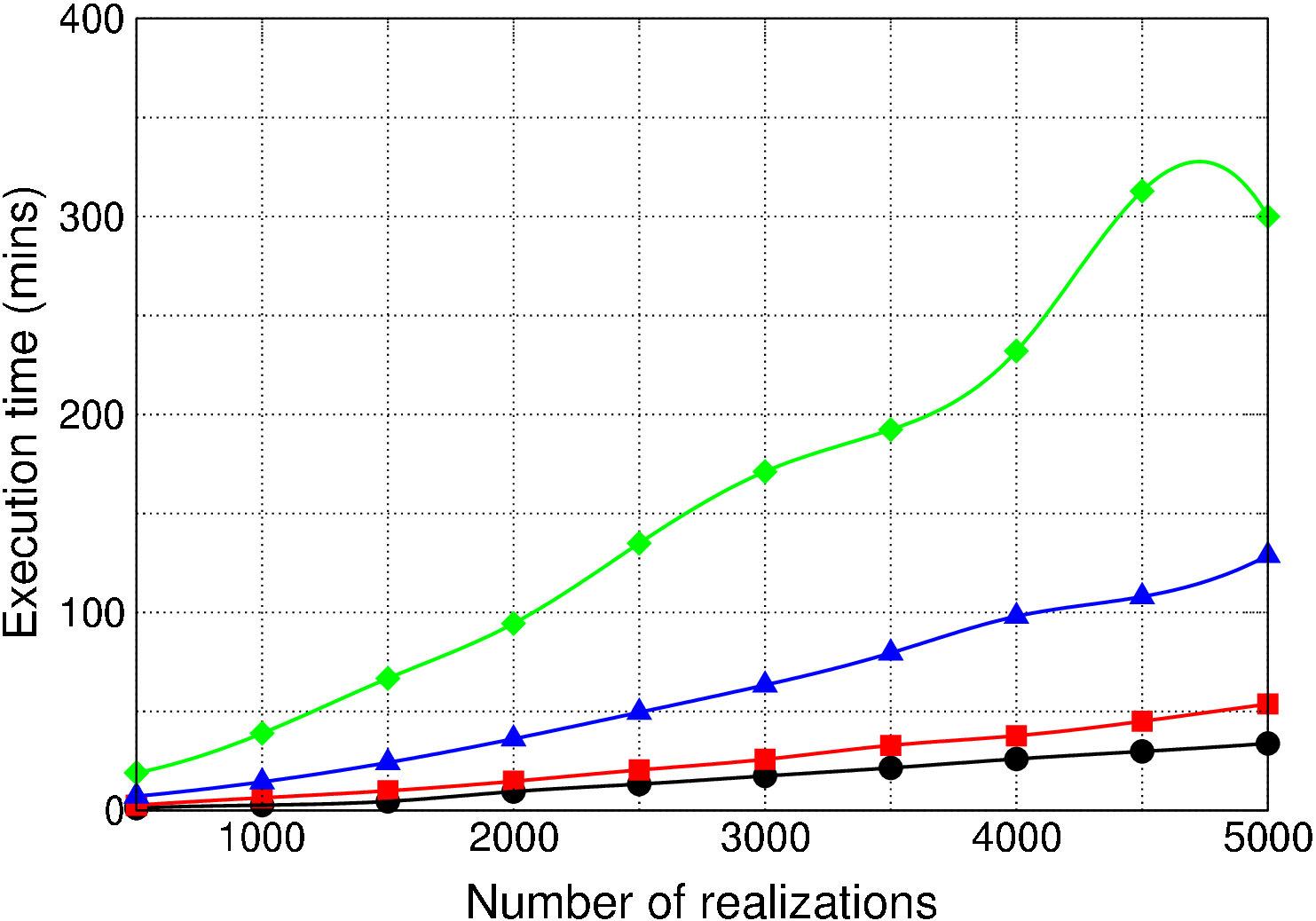}\label{CPU}}\\
\subfigure[]{\includegraphics*[width=0.49\columnwidth]{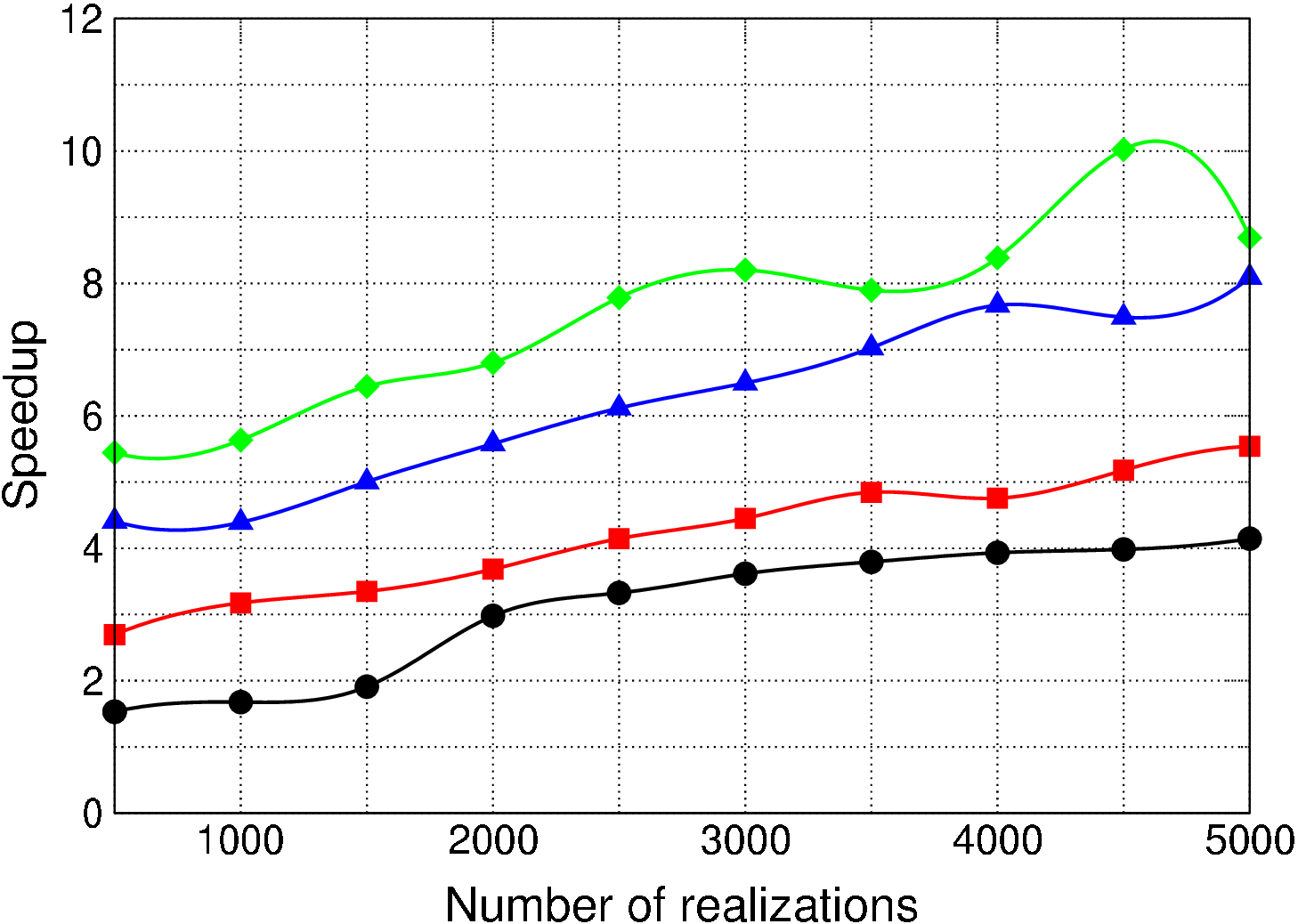}\label{SU}}
\caption{Scaling behavior on Tesla K40 GPU \ref{GPU} and Xeon E3 CPU \ref{CPU} for different sizes of the mesh (from 1600 to 10000 rows), as a function of the number of realizations. The corresponding speedup is illustrated in panel \ref{SU}.
}
\label{fig:Realizations}
\end{figure}

The speedup depends also on the number of realizations considered during parallel execution. About a +3x gain is observed when the number of realizations increases from 500 to 5000 ({\figurename} \ref{fig:Realizations}), irrespectively of the size of the mesh. While the GPU codes scale with the number of realizations (as should be according to the discussion of Sects.\ \ref{RK} and \ref{SERIES}), a performance loss is found for the OPENMP implementation. For the sake of truth, we recall that the OPENMP code was derived from the CUDA code with the strict constraint of adhering as much as possible to it and allowing a fair direct comparison of the computational burden, without introducing any further memory nor algorithmic optimizations. Since the performance loss does not significantly depend on the size of the problem in hand, but only to the number of realizations, this poor behavior can primarily be ascribed to the larger number of calls to the BLAS functions.

\begin{figure}
\centering
\subfigure[]{\includegraphics*[width=0.49\columnwidth]{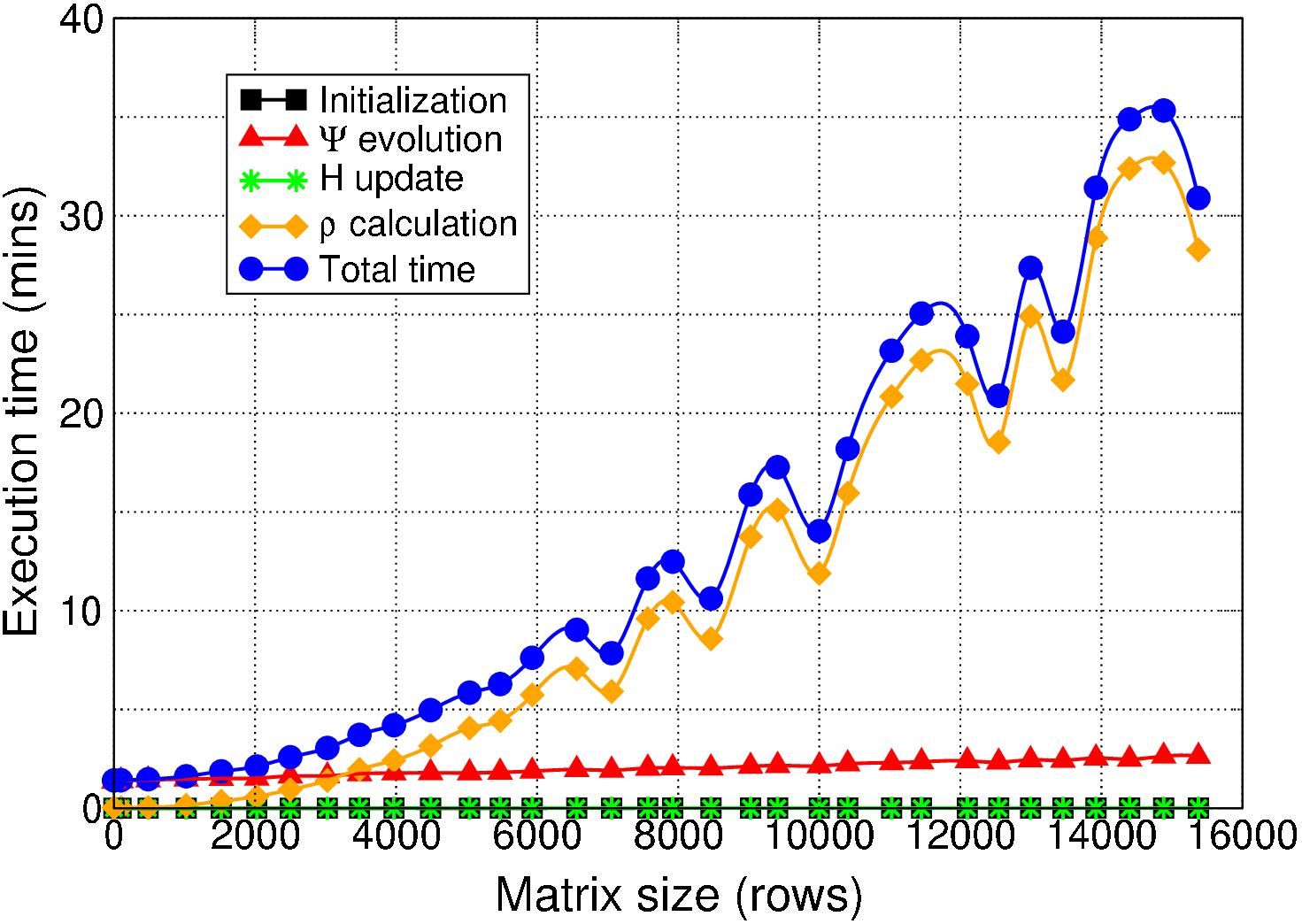}\label{prof_10}}
\hfill
\subfigure[]{\includegraphics*[width=0.49\columnwidth]{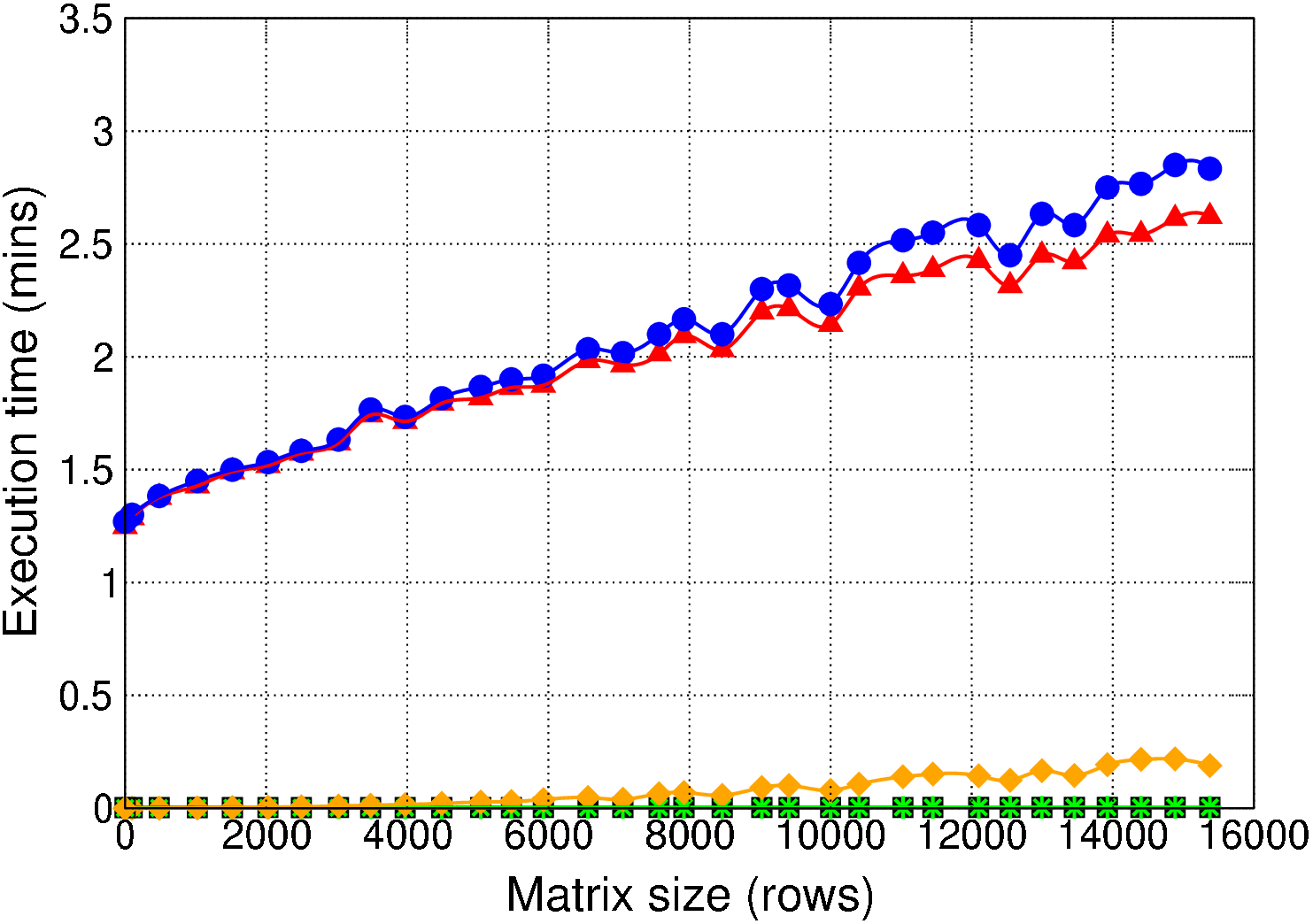}\label{prof_1500}} \\
\hfill
\subfigure[]{\includegraphics*[width=0.49\columnwidth]{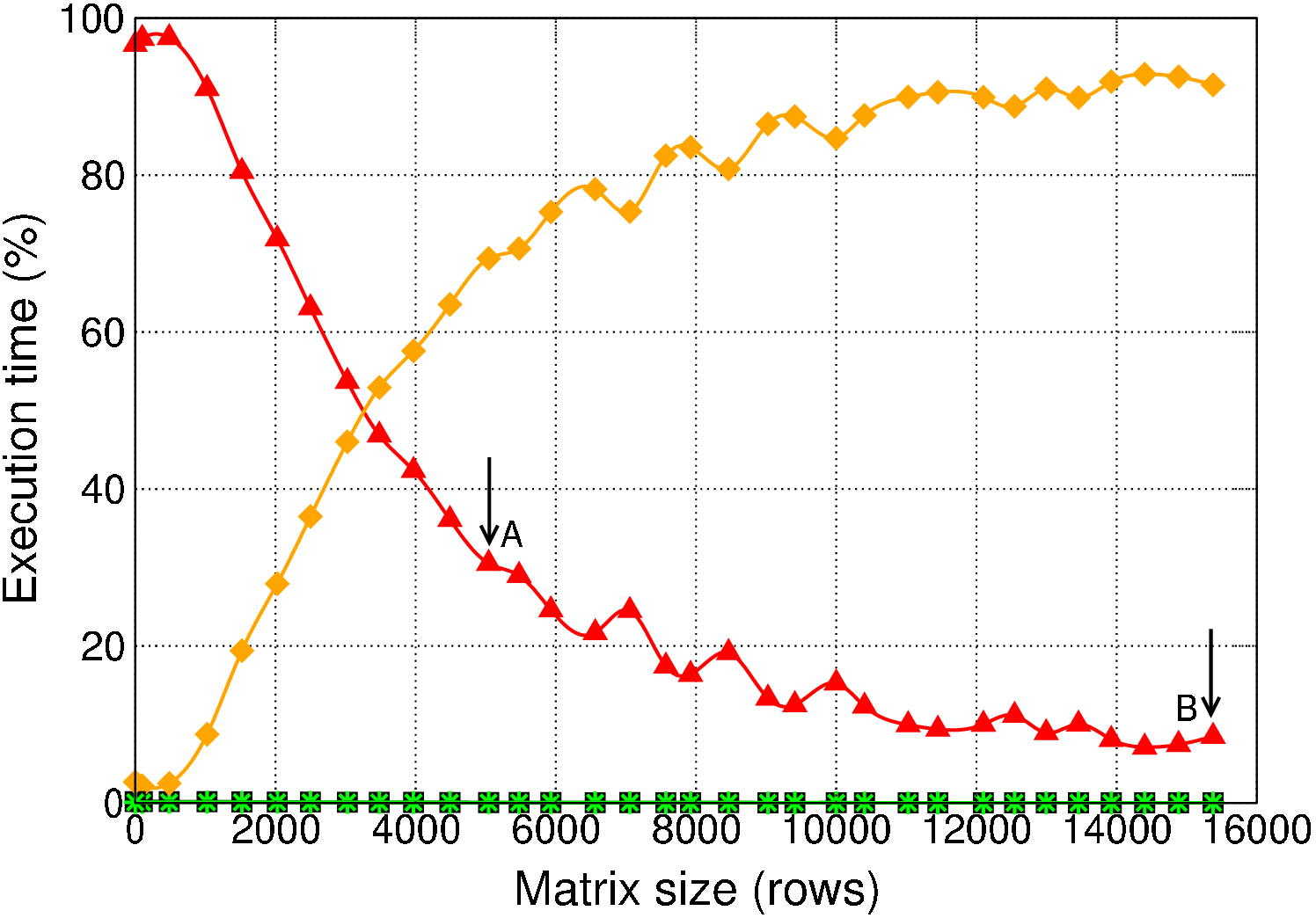}\label{perc_10}}
\hfill
\subfigure[]{\includegraphics*[width=0.49\columnwidth]{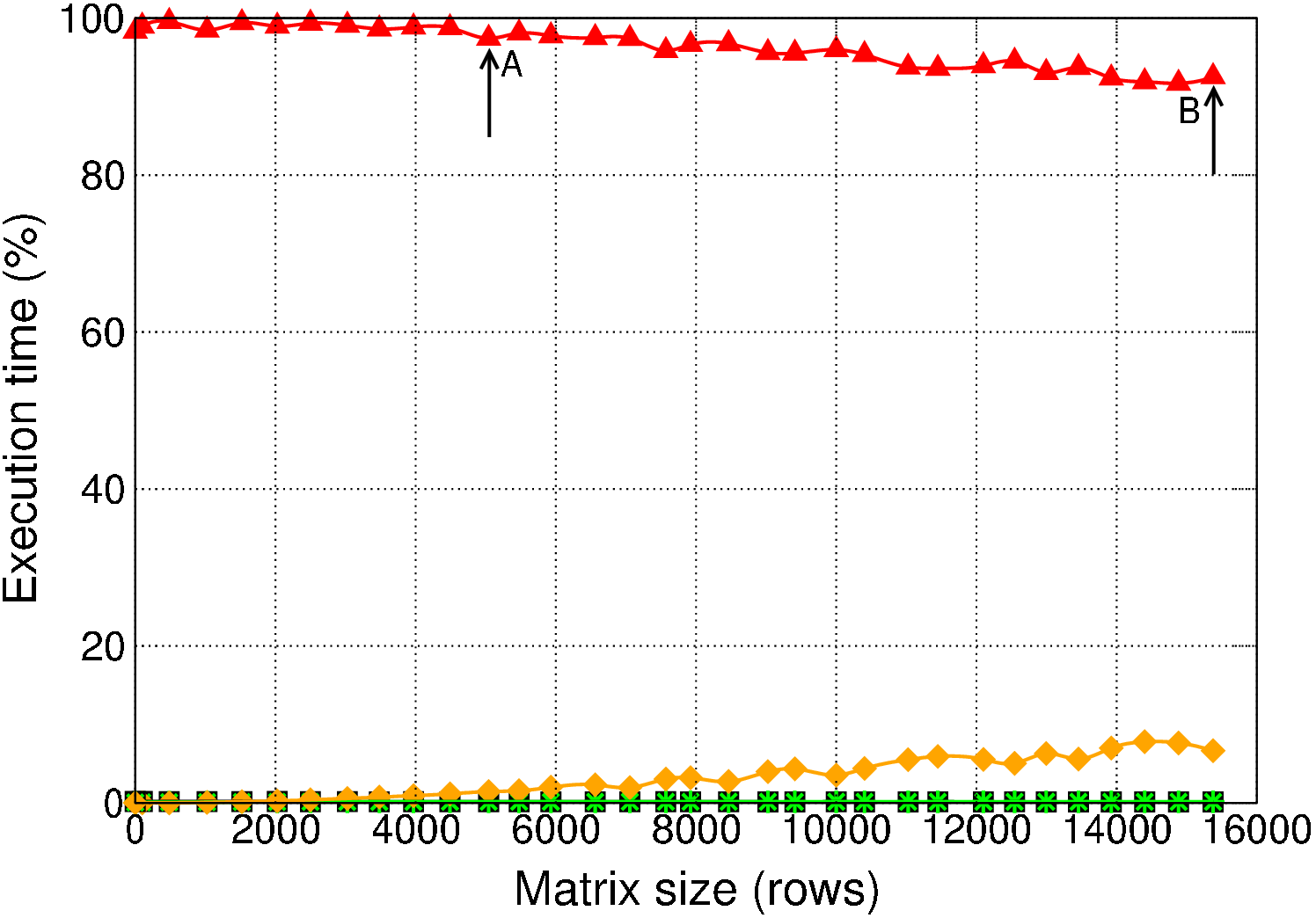}\label{perc_1500}}\\
\subfigure[]{\includegraphics*[width=0.69\columnwidth]{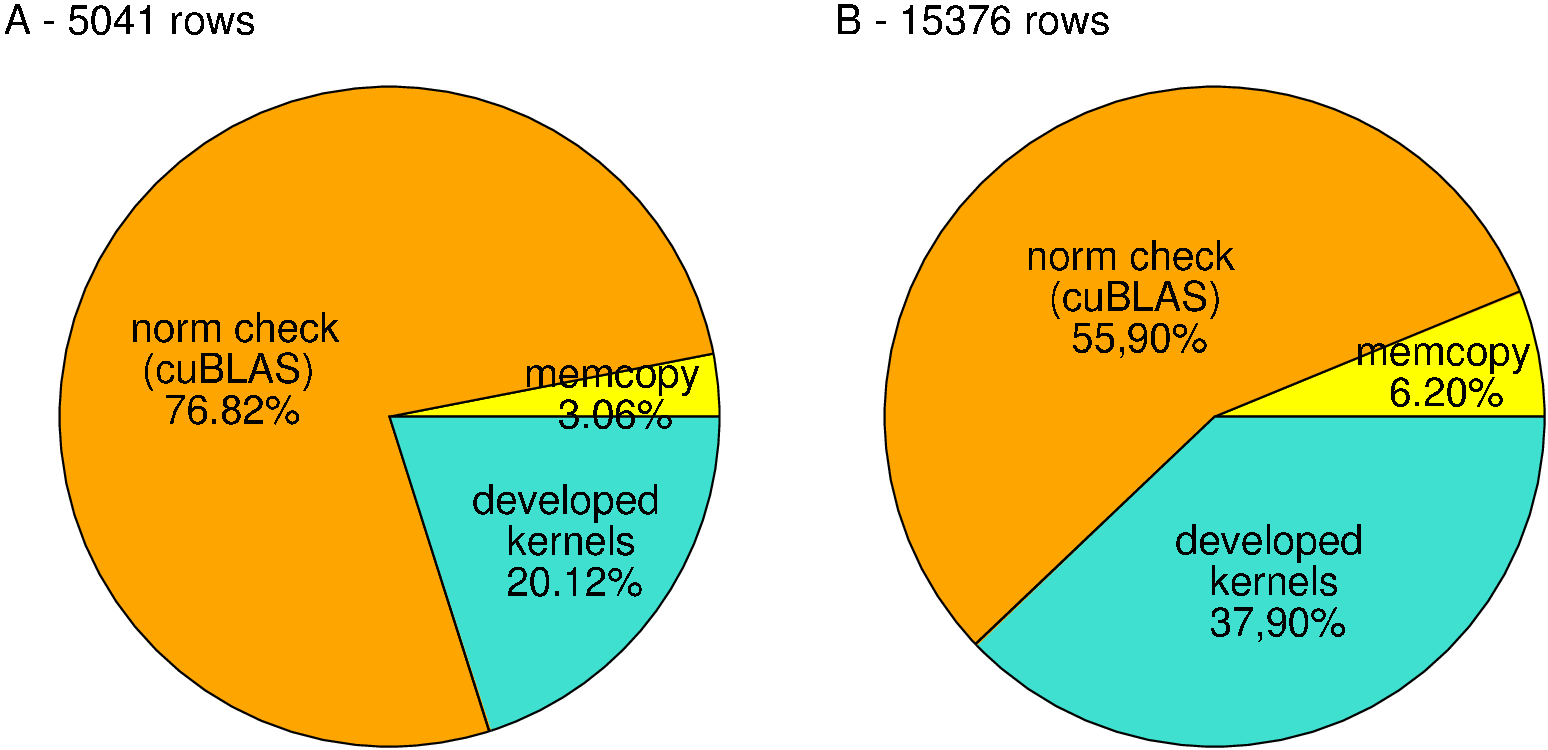}\label{pies}}
\caption{Code profiling and relative weight of the four execution stages for a frequent (\ref{prof_10} and \ref{perc_10}) and for the tiniest (\ref{prof_1500} and \ref{perc_1500}) output generation rate. The pie-charts \ref{pies} show the time required by the sub-components of the wave-function evolution stage for matrix sizes identified by letters A and B. No substantial difference is found between the two cases.
}
\label{fig:profiling}
\end{figure}

The influence of the post-processing rate in GPU execution is even more evident from the shape of the curves of {\figurename} \ref{fig:GPU} that changes from parabolic to linear. Though not immediate at first sight, the same also applies for the curves of {\figurename} \ref{fig:CPU} and is validated by numerical regression. Further information stem from code profiling. We have tracked the execution time of the four stages composing the software for the two opposite cases of very frequent and tiny output generation on the K40 board ({\figurename} \ref{fig:profiling}): the \emph{initialization} and the \emph{Hamiltonian update} stages contribute with a negligible running time (less then 0.3\% in total), while the \emph{wave-function evolution} and the \emph{density-matrix calculation and post-processing} stages largely prevail. 

The running time in case of a very limited output generation is substantially dictated by the \emph{wave-function evolution} stage, which grows linearly with the size of the problem as discussed in Sect.\ \ref{ALGO}. On the contrary, in case of a frequent output generation, the heaviest stage is represented by \emph{density-matrix calculation and post-processing}, whose influence quickly grows up and saturates about 90\% of the total execution time. Going into details, more than the 99.3\% of the time spent for post-processing is required by the \texttt{cublasCher} library function that builds up the average density-matrix $\langle \boldsymbol{\rho}(t) \rangle$. As a consequence, the peaks in panels a, b and c of {\figurename} \ref{fig:SU} are due to outperforming conditions of the cuBLAS library. Also in the \emph{wave-function evolution} stage most of the time is spent in calls to system or library functions (see \figurename \ref{pies}). As a matter of fact, even for large matrices (i.e., $N^m>10000$), only up to approximately one third of the time is dedicated to the series expansion, whereas the remainder is due to device-to-device memcopy and norm evaluation (\texttt{cublasScnrm2} and \texttt{cublasCsscal}). As we already pointed out, memory optimization for speed using, e.g., shared memory on the device, was not a goal of the present work. From the time profiling above, we do not believe it worth the effort: highly-optimized solutions able to cut the execution time of the computational kernels by a factor of 2 or 3 would only bring a very modest benefit around 1 minute or less. To obtain a further significant speedup it is instead mandatory to implement new kernels for linear algebra, other than those provided by the cuBLAS library.

\section{Conclusions}\label{CONC}

The availability of a simulation tool for evolving many-particle CTQWs in a noisy environment represents a crucial prerequisite for the investigation of quantum many-body systems and for the implementation of effective quantum algorithms in realistic situations. In essence, the dynamics of a many-particle state over a noisy lattice can be associated with the solution of a set of stochastic differential equations. However, the need to post-process a large number of data in order to achieve information for any measurable quantity makes the problem much more resource-demanding. In fact, as long as the number of particles and/or the dimensionality of the domain increase, limiting factors such as the memory occupancy and the time required to run the simulations quickly become very challenging issues and determine whether a simulation scheme can or cannot provide results within the available computational power.

Though numerically accurate, the standard Hamiltonian diagonalization method is not feasible even for small systems and alternative numerical solutions must be sought. Among them, we have shown that the 4-th order Runge-Kutta integration  method and the Taylor-series expansion of the evolution operator have a low computational cost and provide reliable data. Moreover, they are highly parallelizable within the SIMT paradigm, and this allows the straightforward, direct implementation on GPUs. 

After developing the codes, we have benchmarked four NVIDIA GPUs and three quad-core Intel CPUs for a 2-particle system over a lattice of increasing dimensions. GPU execution enables significant cuts of the running time of batches of thousands of simulations down to the minute-to-few-hour range. The speedup with respect to OPENMP parallelization stays in the range from 8x to more than 20x, depending on the frequency of post-processing. Our results show that GPU-accelerated codes allow one to overcome concerns about the execution time and make it possible to design simulations involving many particles or large lattices, whose only limit is dictated by the memory available on the device.

\section*{Acknowledgements}

This work has been supported by EU through the
Collaborative Project QuProCS (Grant Agreement 641277) and by UniMI
through the H2020 Transition Grant 15-6-3008000-625 and by UniMoRe through FAR2014.

\end{document}